\documentclass[pra,notitlepage,nofootinbib,twocolumn]{revtex4-2}
\pdfoutput=1
\usepackage[english]{babel}
\usepackage[utf8]{inputenc}
\usepackage[dvipsnames]{xcolor}
\usepackage{amsfonts, amsmath, amssymb, dsfont, graphicx, physics, cancel, multirow}
\usepackage[colorlinks=true,urlcolor=Blue,citecolor=Blue,linkcolor=Blue,anchorcolor=Blue]{hyperref}
\usepackage{tabularx}
\usepackage{bm}
\usepackage{verbatim}
\usepackage{soul}
\usepackage{ulem}

\begin{document}

\title{Filtering variational quantum algorithms for combinatorial optimization}

\author{David Amaro}
\email{david.amaro@cambridgequantum.com}
\affiliation{Cambridge Quantum Computing Limited, SW1P 1BX London, United Kingdom}

\author{Carlo Modica}
\affiliation{Cambridge Quantum Computing Limited, SW1P 1BX London, United Kingdom}

\author{Matthias Rosenkranz}
\affiliation{Cambridge Quantum Computing Limited, SW1P 1BX London, United Kingdom}

\author{Mattia Fiorentini}
\affiliation{Cambridge Quantum Computing Limited, SW1P 1BX London, United Kingdom}

\author{Marcello Benedetti}
\affiliation{Cambridge Quantum Computing Limited, SW1P 1BX London, United Kingdom}

\author{Michael Lubasch}
\affiliation{Cambridge Quantum Computing Limited, SW1P 1BX London, United Kingdom}

\date{\today}

\begin{abstract}
Current gate-based quantum computers have the potential to provide a computational advantage if algorithms use quantum hardware efficiently.
To make combinatorial optimization more efficient, we introduce the Filtering Variational Quantum Eigensolver (F-VQE) which utilizes filtering operators to achieve faster and more reliable convergence to the optimal solution.
Additionally we explore the use of causal cones to reduce the number of qubits required on a quantum computer.
Using random weighted MaxCut problems, we numerically analyze our methods and show that they perform better than the original VQE algorithm and the Quantum Approximate Optimization Algorithm (QAOA).
We also demonstrate the experimental feasibility of our algorithms on a Honeywell trapped-ion quantum processor.
\end{abstract}

\maketitle

\section{Introduction}

Combinatorial optimization tackles problems of practical relevance~\cite{KoVy06}.
Applications include finding the shortest route via several locations for a delivery service, making optimal use of available storage space in logistics, and optimizing a manufacturing supply chain to increase the productivity of a factory.
If quantum algorithms can solve such problems even just slightly faster than classical algorithms, this can have a large impact on various sectors in industry and research.

Variational quantum algorithms are a promising tool to get the most out of the current generation of gate-based quantum processors~\cite{Benedetti_2019, CeEtAl20, Prieto2020, BhEtAl21, Lorenzo2021, Saleem2021}.
These algorithms employ parameterized quantum circuits that can be tailored to hardware constraints such as qubit connectivities and gate fidelities.
In this context, a common approach for combinatorial optimization encodes the optimal solution in the ground state of a classical multi-qubit Hamiltonian~\cite{Kochenberger2014, Lu14, Glover2019}.
Popular variational quantum algorithms such as the Variational Quantum Eigensolver (VQE)~\cite{Peruzzo2014} and the Quantum Approximate Optimization Algorithm (QAOA)~\cite{Farhi2014} attempt to prepare this ground state by searching for the circuit parameters that minimize the energy expectation value of the corresponding quantum state.
VQE imposes no restrictions on the ansatz circuit and has become a powerful method for quantum chemistry~\cite{Moll2018}, condensed matter~\cite{Prieto2020VQE}, and combinatorial optimization~\cite{Saez2018}.
For combinatorial optimization problems, however, it tends to produce sub-optimal solutions~\cite{Diez2021}.
QAOA uses a specific ansatz circuit inspired by adiabatic quantum computation~\cite{FaEtAl01} and the Trotterization of the time evolution corresponding to quantum annealing~\cite{TaHi98}.
Despite its promising properties~\cite{FaHa19, Zhou2020, Moussa2020} and considerable progress with regards to its experimental realization~\cite{HaEtAl21}, in general the QAOA ansatz requires circuit depths that are challenging for current quantum hardware.

\begin{figure}
\centering
\includegraphics[width=0.9\linewidth]{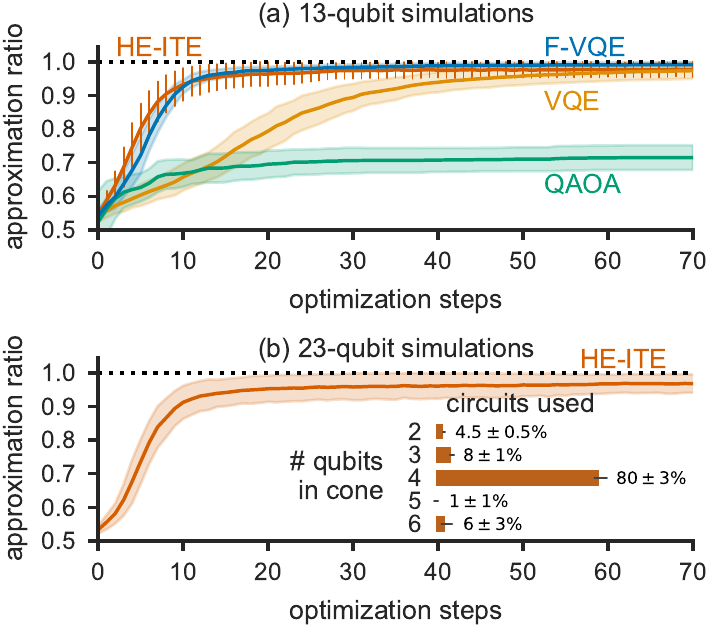}
\caption{\label{fig_main}
Performance of some of our algorithms.
F-VQE employs the inverse filtering operator.
One optimization step corresponds to one time step in HE-ITE and one update of all parameters in F-VQE, VQE, and QAOA.
(a) Average approximation ratio (lines) and standard deviation (error bars for HE-ITE and shaded regions for F-VQE, VQE and QAOA) across 25 random weighted MaxCut problems.
(b) Average approximation ratio (line) and standard deviation (shaded region) across 25 random weighted MaxCut problems.
Inset: causal cone widths -- i.e.\ the actual numbers of qubits required on quantum hardware -- and their average frequency with standard deviation.
}
\end{figure}

In this article, we introduce the Quantum Variational Filtering (QVF) algorithm that optimizes a parameterized quantum circuit to approximate the action of a filtering operator on this circuit.
We also present Filtering VQE (F-VQE) -- a special case of QVF -- which is particularly efficient and similar to VQE.
The main focus of this article is F-VQE which, due to its low quantum hardware requirements, is particularly relevant for current quantum computers.
We consider filtering operators $F \equiv f(\mathcal{H};\tau)$ defined via real-valued functions $f$ of the problem Hamiltonian $\mathcal{H}$ and a parameter $\tau$ in such a way that $f^2(E;\tau)$ strictly decreases with the energy $E$ for any $\tau > 0$.
The parameter $\tau$ plays a role similar to the time step in imaginary time evolution (ITE) and the ITE operator $\exp(-\tau \mathcal{H})$ is one example of a filtering operator considered in this work.
The repeated action of a filtering operator on a quantum state projects out high-energy eigenstates (corresponding to sub-optimal solutions of the combinatorial optimization problem) and increases the overlap with the ground state.
Importantly, QVF and F-VQE have no restrictions on the ansatz circuit and so they can employ the ansatz most suitable for the quantum hardware at hand.
Furthermore we address the question which filtering operators benefit from using causal cones in the optimization, as that can drastically reduce the required number of qubits of the quantum hardware.
Among the filtering operators considered in this article we find that the ITE operator is the best performing one that can be combined with causal cones.
We therefore focus on the combination of ITE with causal cones for which the F-VQE method is equivalent to the hardware-efficient ITE procedure in~\cite{BeFiLu20} (HE-ITE).

We investigate the performance of F-VQE for various filtering operators and of HE-ITE using MaxCut problems on random 3-regular weighted graphs of different sizes.
Finding the optimal solution for this class of MaxCut problems is NP-hard~\cite{Hastad2001, Berman2002} and therefore no classical polynomial-time algorithm is expected to exist that achieves this goal.
Given a weighted graph, the MaxCut problem consists in finding the optimal cut: a separation of the vertices into two disjoint subsets so that the cut cost, i.e.\ the sum of the weights of the edges between the two subsets, is maximum.
The approximation ratio of a cut is defined as the cut cost divided by the cost of the optimal cut.
As shown in Fig.~\ref{fig_main}, F-VQE consistently achieves larger approximation ratios after fewer optimization steps than VQE and QAOA.
Moreover, F-VQE readily runs on actual quantum processors.
The HE-ITE algorithm achieves similar results with a reduced number of qubits and gates.

This article is structured as follows. Section~\ref{s_methods} introduces filtering operators, QVF and F-VQE.
Section~\ref{sec_results} presents the numerical and experimental studies.
We conclude this article and discuss potential next steps in Section~\ref{s_conc_out}.
Technical details are provided in Appendices at the end of this article.

In another publication~\cite{AmEtAl21} we analyse the performance of F-VQE in the context of the job shop scheduling problem.

\section{Methods}
\label{s_methods}

In this Section, we first define filtering operators.
Then we explain the QVF and F-VQE algorithms.
For F-VQE we present a procedure that dynamically updates $\tau$ during the optimization.
Furthermore we address the question which filtering operators can use causal cones in F-VQE.

\subsection{Filtering operators}
\label{s_dh_fo}

Given a $n$-qubit Hamiltonian $\mathcal{H}$, we define a \emph{filtering operator} $F \equiv f(\mathcal{H};\tau)$ via a real-valued function $f(E;\tau)$ of the energy $E$ and a parameter $\tau > 0$.
We require that the function $f^2(E;\tau)$ is strictly decreasing on the interval given by the complete spectrum of the Hamiltonian $E \in [E_{\min}, E_{\max}]$.
Filtering operators are Hermitian and commute with the Hamiltonian by definition.

\begin{figure}[t]
\centering
\includegraphics[width=0.9\linewidth]{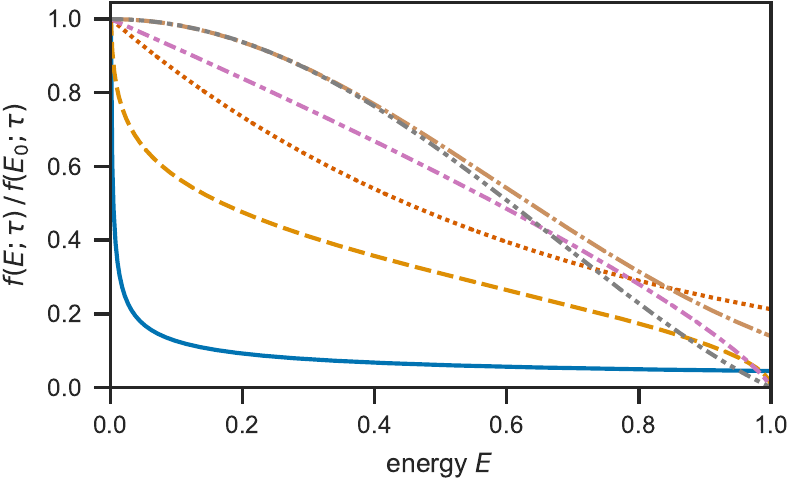}
\vspace{0.4cm}\\
\begin{tabular}{c|ccc}
Filter & $f(\mathcal{H};\tau)$ & $\tau_1$ & Line\\ 
\hline 
Inverse & $\mathcal{H}^{-\tau}$ & $0.45\pm0.20$ & \parbox[c]{1em}{\includegraphics[width=0.2in]{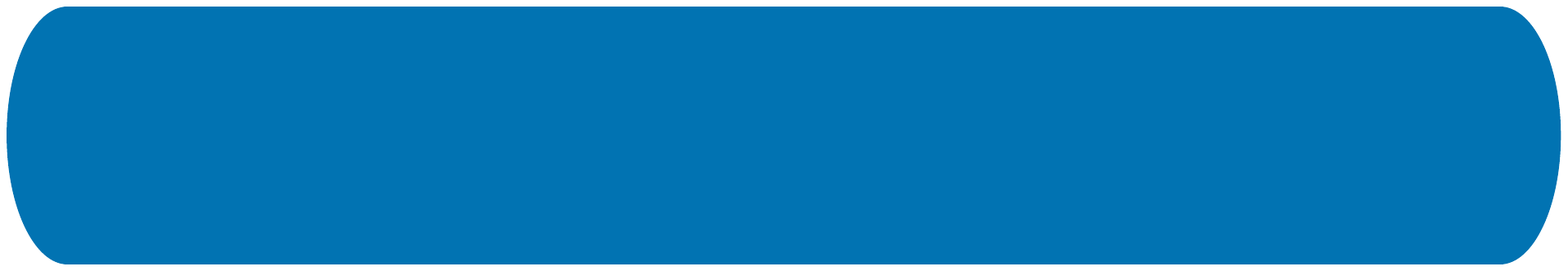}}\\[-0.2cm]
Logarithm & $(-\log(\mathcal{H}))^{\tau}$ & $0.510\pm0.091$ & \parbox[c]{1em}{\includegraphics[width=0.2in]{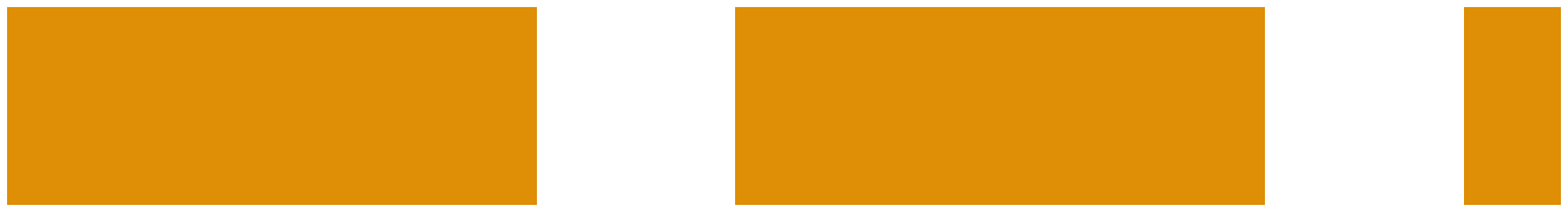}}\\[-0.2cm]
Exponential & $e^{-\tau \mathcal{H}}$ & $1.55\pm0.28$ & \parbox[c]{1em}{\includegraphics[width=0.2in]{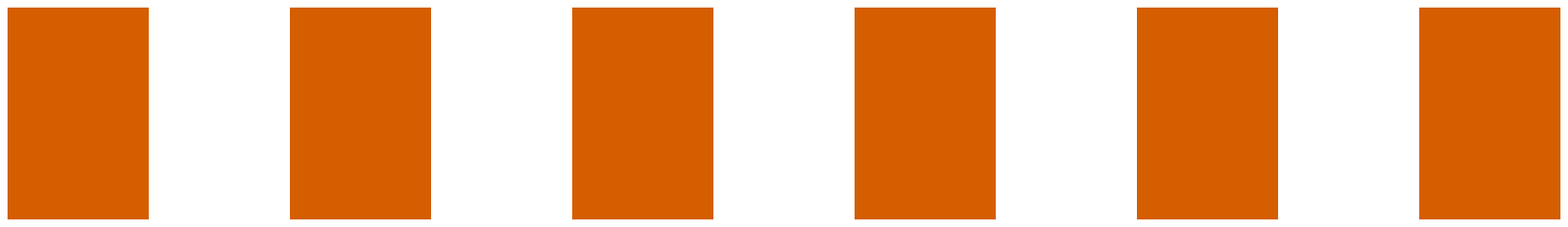}}\\[-0.2cm]
Power & $(\mathds{1} - \mathcal{H})^\tau$ & $0.79\pm0.17$ & \parbox[c]{1em}{\includegraphics[width=0.2in]{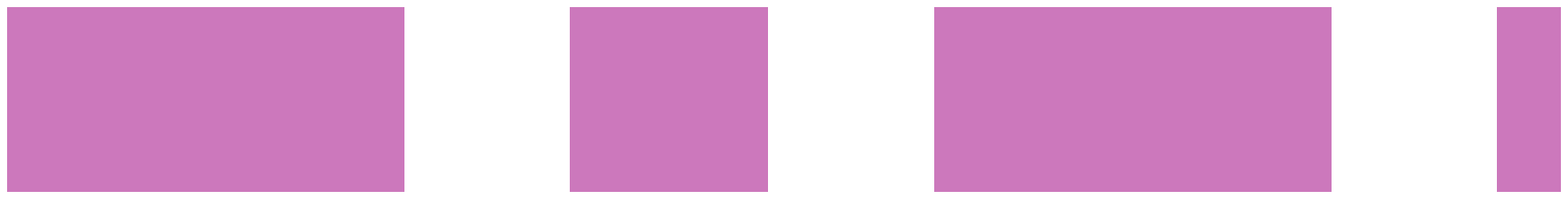}}\\[-0.2cm]
Cosine & $\cos^\tau(\mathcal{H})$ & $3.20\pm0.74$ & \parbox[c]{1em}{\includegraphics[width=0.2in]{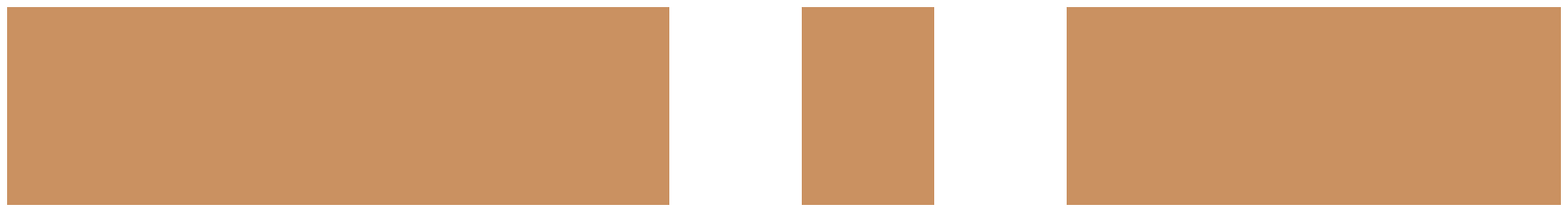}}\\[-0.2cm]
Chebyshev & Eq.~(\ref{eq_che}) &$4.52\pm0.71\approx 5$ & \parbox[c]{1em}{\includegraphics[width=0.2in]{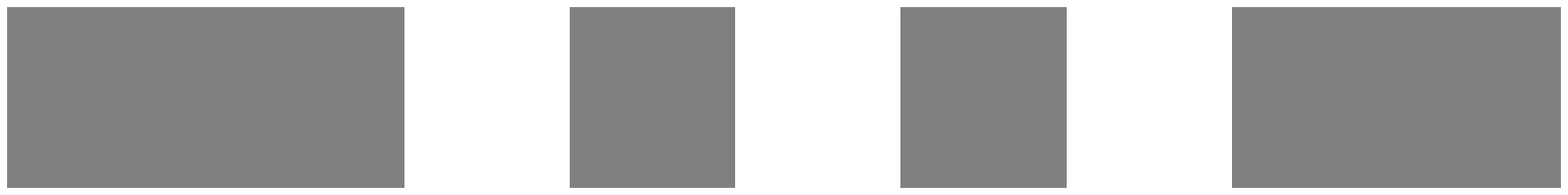}}
\end{tabular}
\caption{\label{f_filter_functions}
Filtering operators used in this work.
The value of $f(E;\tau) / f(E_0;\tau)$ is plotted against the energy range $E \in [E_0, 1]$ for ground state energy $E_0 = 0.001$.
The value of $\tau$ used in the figure is the first value $\tau_1$, selected at the first optimization step, averaged across the 25 MaxCut instances for 13 qubits.
Values after $\pm$ indicate the standard deviation.
For the Chebyshev filter, the value used in the figure was rounded to $5$.
}
\end{figure}

When a quantum state $\ket{\psi}$ is sampled in the eigenbasis $\{\ket{\lambda_x}:\:x=0,\,1,\ldots,\,2^n-1\}$ of a Hamiltonian, then the probability distribution of eigenvectors is given by $P_\psi(\lambda_x) = |\langle\psi|\lambda_x\rangle|^2$.
The application of the filtering operator on the quantum state produces a new quantum state $\ket{F\psi} = F\ket{\psi}/\sqrt{\langle F^2\rangle_\psi}$ which generates a probability distribution that depends on the energy $E_x$ of each eigenstate $\ket{\lambda_x}$:
\begin{equation}\label{eq_distr}
P_{F\psi}(\lambda_x) = \frac{f^2(E_x;\tau)}{\langle F^2 \rangle_\psi}P_{\psi}(\lambda_x) .
\end{equation}
For non-eigenstates $\ket{\psi}$ the action of the filtering operator increases the probability of all overlapping eigenstates $\ket{\lambda_x}$ (where $P_{\psi}(\lambda_x) > 0$) for which $f^2(E_{x};\tau) > \langle F^2\rangle_\psi$ and decreases the probability of all overlapping eigenstates $\ket{\lambda_{y}}$ for which $f^2(E_{y};\tau) < \langle F^2\rangle_\psi$.
Since $f^2(E;\tau)$ is strictly decreasing as a function of $E$, such eigenstates exist for any non-eigenstate $\ket{\psi}$.
Hence, under the application of the filtering operator, the probability of sampling eigenstates with low energy increases and the probability of sampling eigenstates with high energy decreases.
This naturally leads to a reduction of the average energy.
In particular, if the ground state $\ket{\lambda_0}$ has a finite overlap ($P_{\psi}(\lambda_0) > 0$) with the initial state, the probability of sampling it increases with every application of the filtering operator.
After sufficiently many applications the ground state is produced.

Some filtering function definitions are given in Fig.~\ref{f_filter_functions}.
The Chebyshev filtering operator approximates the Dirac delta operator $\delta(\mathcal{H})$ using the following expansion in terms of Chebyshev polynomials up to order $\tau$~\cite{Weiss2006}:
\begin{align}
\delta(\mathcal{H}) \approx & f(\mathcal{H};\tau) = \sum_{r = 0}^{\lfloor \tau / 2 \rfloor} (-1)^{r} \frac{2 - \delta_{r, 0}}{\pi} g_{2 r}^{(\tau)} T_{2 r}(\mathcal{H}) \label{eq_che}\\
g_{s}^{(\tau)} & = \frac{(\tau - s + 1) \cos{\frac{\pi s}{\tau + 1}} + \sin{\frac{\pi s}{\tau + 1}} \cot{\frac{\pi}{\tau + 1}}}{\tau + 1} .
\end{align}
Here $\delta_{r, s}$ is the Kronecker delta and the Chebyshev polynomials are defined via the recursive formula $T_{s+1}(x) = 2 x T_{s}(x) - T_{s-1}(x)$ with $T_{0}(x) = 1,\, T_{1}(x) = x$.

The parameter $\tau$ in the filtering operator definition is inspired by the time step parameter of imaginary time evolution.
In fact, for the exponential filtering operator in  Fig.~\ref{f_filter_functions} $\tau$ is precisely the imaginary time step.
This parameter interpolates the action of the filtering operator between two limits.
For vanishing values of $\tau \to 0$ the filtering operator becomes the identity operator and for sufficiently large values of $\tau \to \infty$ the filtering operator becomes a projector onto the ground state.

For our selection of filters we took inspiration from several works.
The inverse filter is inspired by the inverse iteration procedure which is a common ingredient in numerical routines that calculate eigenvalues and eigenvectors of matrices~\cite{TrBa97, NoLuJe13, NoEtAl17}.
The cosine filter was previously used in non-variational algorithms to achieve faster ground state preparation~\cite{ge2018faster} and to analyze finite energy intervals~\cite{LuBaCi20}.
The Chebyshev filter considered here was also employed in powerful tensor network algorithms for the study of thermalization~\cite{YaEtAl20, BaHuCi20, CaCiBa21}.

\subsection{Quantum Variational Filtering (QVF)} \label{sec:IIB}

The QVF algorithm approximates the repeated action of a filtering operator on some initial quantum state by successively optimizing the variational parameters of a parameterized quantum circuit.
The algorithm starts at optimization step $t = 0$ by preparing an initial state $\ket{\psi_0}$ that has finite overlap with the ground state: $P_{\psi_0}(\lambda_0) > 0$.
Then the algorithm proceeds iteratively and in each optimization step $t \geq 1$ approximates the state $\ket{F_t\psi_{t-1}}$ -- that results from exactly applying the filtering operator $F_t$ to the state $\ket{\psi_{t-1}}$ -- by a state $\ket{\psi_t}$.
The subscript $t$ in $F_t$ indicates that the filtering operator can change at each optimization step.
The algorithm stops after an initially chosen number of optimization steps.

In order to approximate the application of the filtering operator, we prepare a parameterized quantum circuit ansatz $\ket{\psi(\bm{\theta})}$ that depends on a vector of $m$ parameters $\bm{\theta} = (\theta_1,\,\ldots,\,\theta_m)$.
At optimization step $t$ we search for the parameters that minimize the Euclidean distance between the parameterized quantum state and $\ket{F_t\psi_{t-1}}$:
\begin{equation}\label{eq_cost_fun}
\begin{split}
\mathcal{C}_t(\bm{\theta}) & = \frac{1}{2}\Vert \ket{ \psi(\bm{\theta}) } - \ket{F_t \psi_{t-1}} \Vert^2\\
& = 1 - \frac{ \Re \bra{\psi_{t-1}} F_t \ket{\psi(\bm{\theta})} }{ \sqrt{ \expval{ F_t^2 }_{\psi_{t-1}}}} .
\end{split}
\end{equation}
The final vector of parameters obtained at the end of the minimization of Eq.~\eqref{eq_cost_fun} defines the quantum state $\ket{\psi_t} \equiv \ket{\psi(\bm{\theta}_t)}$ at optimization step $t$.
The cost function in Eq.~\eqref{eq_cost_fun} can be minimized with the help of the Hadamard test, which needs one additional ancilla qubit and several additional controlled operations, as we explain in Appendix~\ref{app_hadamard}.

\subsection{Filtering VQE (F-VQE)}
\label{subsec:F-VQE}

To avoid the additional quantum resources required by the Hadamard test in QVF, in the following we develop F-VQE.
The F-VQE algorithm uses a specific gradient-based procedure that requires essentially the same circuits as VQE.

The partial derivative of the cost function in Eq.~\eqref{eq_cost_fun} with respect to one parameter $\theta_j$ is derived in Appendix~\ref{app_an_grad_fvqe}:
\begin{equation}\label{eq_grad}
\dfrac{\partial \mathcal{C}_t(\bm{\theta})}{\partial \theta_j} = - \frac{ \Re \bra{\psi_{t-1}} F_t \ket{\psi(\bm{\theta} + \pi \bm{e}_j)}}{2\sqrt{\langle F_t^2 \rangle_{\psi_{t-1}}}} .
\end{equation}
Here the state $\ket{\psi(\bm{\theta} + \pi \bm{e}_j)}$ is produced by the same ansatz circuit except that the vector of angles is shifted by an amount $\pi$ along the direction $\bm{e}_j$ of parameter $\theta_j$.
If the gradient is evaluated at the current vector of parameters $\bm{\theta}_{t-1}$, then the parameter-shift rule~\cite{Mitarai2018, Schuld2019} yields:
\begin{equation}\label{eq_grad_simp}
\left.\dfrac{\partial \mathcal{C}_t(\bm{\theta})}{\partial \theta_j}\right\vert_{\bm{\theta}_{t-1}} = - \frac{ \langle F_t \rangle_{\psi_{t-1}^{j+}} - \langle F_t \rangle_{\psi_{t-1}^{j-}}}{4 \sqrt{\langle F_t^2 \rangle_{\psi_{t-1}}}} .
\end{equation}
Here the three circuits $\ket{\psi_{t-1}}$ and $\ket{\psi_{t-1}^{j\pm}} \equiv \ket{\psi(\bm{\theta}_{t-1} \pm \tfrac{\pi}{2} \bm{e}_j)}$ are generated by the ansatz with different parameter vectors.
Note that the expectation value in the denominator is the same for all partial derivatives at fixed $t$.

The F-VQE algorithm takes advantage of this favorable case as follows.
At optimization step $t$, F-VQE performs a \emph{single} gradient-descent update:
\begin{equation}
\bm{\theta}_t = \bm{\theta}_{t-1} - \eta \sum_{j = 1}^{m} \left.\dfrac{\partial \mathcal{C}_t(\bm{\theta})}{\partial \theta_j}\right|_{\bm{\theta}_{t-1}} \bm{e}_j ,
\end{equation}
where $\eta > 0$ is the learning rate.
Then F-VQE moves on to the next cost function $\mathcal{C}_{t+1}(\bm{\theta})$ and proceeds identically.
For each optimization step this algorithm requires the evaluation of $2 m + 1$ circuits.

The expectation value $\langle F_t \rangle_\psi$ of filtering operators can be efficiently evaluated by sampling the quantum state in the Hamiltonian eigenbasis.
If each eigenstate $\ket{\lambda_x}$ is sampled $M_x$ times from a total of $M$ samples, the filtering operator expectation value can be approximated via the Monte Carlo estimator $f(E;\tau)$ as:
\begin{equation}\label{eq_exp_val}
\langle F_t \rangle_\psi \approx \frac{1}{M} \sum_{x} M_x f(E_{x};\tau) .
\end{equation}

In this article, we represent combinatorial optimization problems by diagonal QUBO (Quadratic Unconstrained Binary Optimization) Hamiltonians~\cite{Kochenberger2014, Lu14, Glover2019} for which the eigenbasis is the computational basis and energies can be efficiently computed.
Therefore the expectation value of a filtering operator can be approximated by sampling the quantum state in the computational basis.

At each optimization step $t$ the samples used to compute $\langle F_t^2 \rangle_{\psi_{t-1}}$ in Eq.~\eqref{eq_grad_simp} are also used to compute the average energy $\langle \mathcal{H} \rangle_{\psi_{t-1}}$.
As $t$ increases, the average energy is expected to decrease and the probability of sampling the ground state is expected to increase.
Thus F-VQE provides the average energy and a growing chance of sampling a low energy eigenstate or even the ground state at no extra cost during the optimization.

The gradient in F-VQE is equivalent to the one in VQE under certain assumptions.
We derive the VQE gradient in Appendix~\ref{app_an_grad_vqe}.
If the Hamiltonian in the VQE gradient of Eq.~\eqref{eq_vqe_an_grad} is replaced by $-F_t$, the new VQE gradient evaluated at the point $\ket{\psi(\bm{\theta})} = \ket{\psi_{t-1}}$ coincides with the F-VQE gradient in Eq.~\eqref{eq_grad_simp} up to a positive multiplicative factor.
However, we emphasize that the corresponding VQE cost function $-\bra{\psi(\bm{\theta})} F_t \ket{\psi(\bm{\theta})}$ is different from the F-VQE cost function in Eq.~\eqref{eq_cost_fun} where the dependence on the parameters is of the form $-\text{Re}\bra{\psi_{t-1}} F_t \ket{\psi(\bm{\theta})}$. 
We note that the F-VQE gradient in Eq.~\eqref{eq_grad}, in general, coincides only at the point $\ket{\psi(\bm{\theta})} = \ket{\psi_{t-1}}$ with the gradient of the modified VQE cost function.
We can also see in Eqs.~\eqref{eq_second_der} and \eqref{eq_second_der_vqe} that the second derivatives do not coincide, not even at that point $\ket{\psi(\bm{\theta})} = \ket{\psi_{t-1}}$.
Therefore both algorithms explore parameter landscapes with different curvatures.

\subsection{Adapting \texorpdfstring{$\tau$}{tau}}
\label{s_tau_sch}

Both the cost function in Eq.~\eqref{eq_cost_fun} and its gradient in Eq.~\eqref{eq_grad_simp} depend on the parameter $\tau$ via the expectation value of the filtering operator in Eq.~\eqref{eq_exp_val}.
We dynamically adapt $\tau$ to keep the gradient norm as close as possible to some desired large and fixed value at every optimization step.
This can prevent the gradient from vanishing and enable us to determine its value more accurately with a fixed number of measurements.

\begin{figure}[t]
\centering
\includegraphics[width=.9\linewidth]{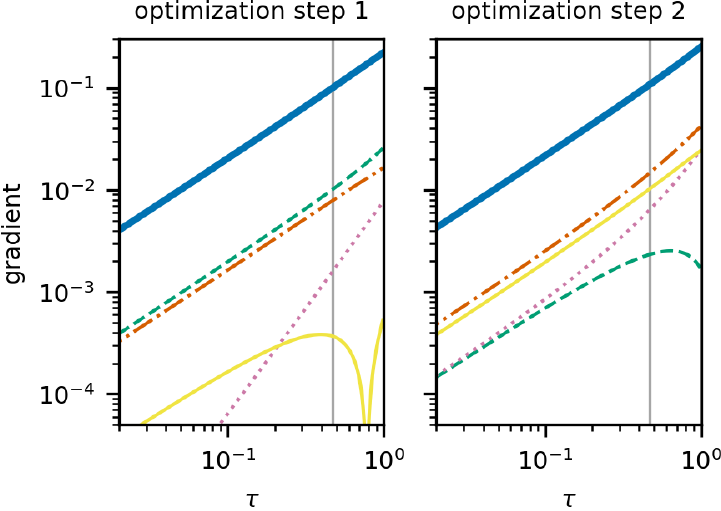}
\caption{\label{f_grad_vs_tau}
Gradients as a function of the parameter $\tau$.
For one 13-qubit MaxCut problem we plot the gradient norm (blue continuous) and the absolute value of partial derivatives in Eq.~\eqref{eq_grad_simp} with respect to four parameters in the ansatz circuit shown in Fig.~\ref{fig_ansatz}(a).
Partial derivatives correspond to the first parameter on qubit 1 (green dashed), the last parameter on qubit 1 (magenta dotted), the first parameter on qubit 13 (yellow continuous), and the last parameter on qubit 13 (red dashed-dotted).
The value of $\tau_t$ selected at these optimization steps is marked with a gray vertical line.
}
\end{figure}

In F-VQE we employ the following heuristic to dynamically adapt $\tau$.
At each optimization step the dependence $g(\tau) \equiv \Vert\bm{\nabla}\left.\mathcal{C}_t(\tau)\right|_{\bm{\theta}_{t-1}}\Vert $ between the gradient norm and $\tau$ is used to select the value $\tau_t$ that returns a gradient norm as close as possible below a certain threshold $g_c > 0$.
For each optimization step such value $\tau_t$ is obtained by solving the implicit equation $g(\tau_t) = g_c$.
Note that $g(0) = 0$ since for $\tau = 0$ the filtering operator becomes the identity operator and the gradient norm vanishes.
For large values of $\tau$ the gradient norm saturates at a finite value that is determined by the overlap of the gradient circuits with the ground state.
Taking this into account, we select $\tau_t$ in the following way.
We evaluate the gradient norm for increasing values of $\tau$ until either (i) an upper bound $\tau_u > \tau_t$ is found such that $g(\tau_u) > g_c$ or (ii) $g(\tau)$ converges to a constant.
In the first case (i) we search for $\tau_t$ in the range $[0, \tau_u]$ up to a certain precision.
In the second case (ii) we select from the tried values the one that provided a gradient norm closest below the threshold.
Note that for the Chebyshev filtering operator only positive integer values of $\tau$ are allowed.

We emphasize that our heuristic is different from a simple re-scaling of the gradient by a constant.
As shown in Fig.~\ref{f_grad_vs_tau} each partial derivative changes non-trivially as a function of $\tau$.
Moreover, in the simulations we observe that the gradient norm has a consistent dependence on $\tau$ across different optimization steps and problems.

\subsection{Causal cones}
\label{subsec:CausalCones}

The F-VQE algorithm is based on expectation value computations of filtering operators and it is natural to ask whether this method can benefit from using causal cones.
The causal cone of an observable is the quantum circuit composed of only those qubits and gates in the ansatz circuit that have an actual effect on the expectation value.
In general, causal cones allow us to simplify the computation of expectation values for local observables.
The simplification follows from the fact that outside the causal cone of a local operator unitary gates cancel with their adjoints.
Causal cones are a crucial ingredient in various tensor network methods, e.g.\ based on the multiscale entanglement renormalization ansatz~\cite{Vi08}.
We have previously used them to make variational quantum algorithms more hardware-efficient for the simulation of the time evolution of quantum many-body systems~\cite{BeFiLu20}.

Figures~\ref{fig_ansatz}(b) and (c) show the causal cones for observables with support on two neighboring qubits and two non-neighboring qubits, respectively.
Only the qubits and gates inside the causal cone need to be prepared experimentally.
Note that for observables with support on distant qubits, the causal cone splits into two separable causal cones that can be independently realized in hardware.
Therefore causal cones can reduce the required number of gates and qubits when the observables have small support.

Inspecting the filtering operators in Fig.~\ref{f_filter_functions}, we see that the exponential, power, and Chebyshev filters can make use of causal cones.
The exponential filter $\exp(-\tau \mathcal{H})$ is equivalent to a product of 2-local terms that can be processed independently if additional approximations are made~\cite{BeFiLu20}.
The power filter $(1-\mathcal{H})^{\tau}$ for integer values of $\tau$ is equivalent to a sum of at most $2\tau$-local terms so that the entire expectation value can be determined from the sum of simpler expectation values.
Similarly the expectation value of the Chebyshev filter can be calculated using the sum of expectation values of at most $2\tau$-local observables, as the Chebyshev filtering operator is a polynomial in $\mathcal{H}$ of degree $\tau$.

\section{Results} 
\label{sec_results}

In this Section, we describe the MaxCut Hamiltonians, parameterized quantum circuits, as well as the simulation and experimental settings that are being considered in this article.
The settings are summarized in Tab.~\ref{t_settings}.
Then we analyze the performance of the algorithms.
Simulations and the experiment on the Honeywell H1 trapped-ion quantum processor~\cite{Pino2021} are compiled by TKET~\cite{Sivarajah2020}.

\subsection{Weighted MaxCut Hamiltonians}
\label{subsec:problem_hamiltonian}

We use 25 random MaxCut instances for each problem size of $n \in \{5,\, 7,\, 9,\, 11,\, 13,\, 23\}$ qubits.
These problem sizes $n$ correspond to graphs with $N = n+1$ vertices.
Each instance is defined on a random 3-regular weighted simple undirected and connected graph $\mathcal{G}(\mathcal{V}, \mathcal{E}, \mathcal{W})$ where $\mathcal{V}=\{1,\,2,\,\ldots,\,N\}$ is the set of vertices, $\mathcal{E} \subset \mathcal{V} \times \mathcal{V}$ is the set of edges between different vertices, and $\mathcal{W} = \{w_e \in [0,1]:\: e \in \mathcal{E}\}$ is the set of random weights uniformly distributed in the range $[0,1]$ for all edges.
A cut is represented by variables $z_{v}=\{+1, -1\}$, with $v \in \mathcal{V}$, that are $+1$ for the vertices in one subset of the cut and $-1$ for the vertices in the other subset of the cut.
This formulation has the obvious symmetry of swapping labels $+1,\, -1$ for the two subsets.
We break this symmetry by assigning $+1$ to the last vertex $v = N$ and thereby reduce the number of variables to $n = N - 1$.
Then the MaxCut problem consists in solving the optimization problem $(z^*_1, z^*_2, \ldots, z^*_{n}) = \operatorname*{argmax}_{(z_1, z_2, \ldots, z_{n})}C(z_1, z_2, \ldots, z_{n})$, with cost function
\begin{equation} \label{eq_maxcut}
\begin{split}
    C(z_1, z_2, \ldots, z_{n}) &= \sum_{e = \{u, N\} \in \mathcal{E}} w_e \frac{1 - z_u}{2} \\
    &+ \sum_{e = \{u, v \neq N\} \in \mathcal{E}} w_e \frac{1 - z_u z_v}{2} .
\end{split}
\end{equation}

The Hamiltonian formulation of the MaxCut problem is obtained via any real coefficients $a$ and $b > 0$ as
\begin{equation}\label{eq:ham_def}
 \mathcal{H} = a \mathds{1} - b C(Z_1, Z_2, \ldots, Z_n) ,
\end{equation}
where each variable $z_u$ in the MaxCut cost function is replaced by the Pauli operator $Z_u$ acting on qubit $u \in \{1, \ldots, n\}$.
A ground state of $\mathcal{H}$ in Eq.~\eqref{eq:ham_def} is the computational state $\ket{\lambda_0} = \bigotimes_{v=1}^n \ket{(1-z^*_v)/2}$.
The \textit{approximation ratio} of a $n$-qubit quantum state $\ket{\psi}$ is defined as
\begin{equation}
 \langle \alpha \rangle_\psi = \frac{\langle C(Z_1, Z_2, \ldots, Z_n)\rangle_\psi}{\operatorname*{max}_{(z_1, z_2, \ldots, z_{n})}C(z_1, z_2, \ldots, z_{n})} .
\end{equation}

Before we apply filtering operators to MaxCut Hamiltonians, we re-scale the energy range to $[0, 1]$ using the coefficients $a$ and $b$ in Eq.~\eqref{eq:ham_def}.
To achieve this, we compute lower and upper bounds of the MaxCut cost function in Eq.~\eqref{eq_maxcut}.
We choose the upper bound to be the optimum cost of the semidefinite programming (SDP) relaxation of the MaxCut problem~\cite{Helmberg2000}.
We fix the lower bound to the minimum cost $0$, which corresponds to the trivial cut $z_u = +1$ for all $u \in \{1, \ldots, n\}$.

\subsection{Setup} 
\label{subsec:setup}

\begin{table}
\begin{tabular}{l|cccc}
(a) algorithm & F-VQE & HE-ITE & VQE & QAOA \\
\hline
cost function & Eq.~\eqref{eq_cost_fun} & \cite{BeFiLu20} & $\langle \mathcal{H}\rangle_{\psi(\bm{\theta})}$ & $\langle \mathcal{H} \rangle_{\psi(\bm{\gamma}, \bm{\beta})}$ \\
ansatz circuit & Fig.~\ref{fig_ansatz}(a) & Fig.~\ref{fig_ansatz}(a) & Fig.~\ref{fig_ansatz}(a) & Eq.~(\ref{eq_qaoa_ansatz}) \\
initial param. & $\ket{+}^{\otimes n}$ & $\ket{+}^{\otimes n}$ & $\ket{+}^{\otimes n}$ & random \\
adaptive $\tau$ & yes, $g_c = 0.1$ & no, fix $1.0$ & - & - \\ 
learning rate $\eta$ & inv.\ Hess.\ d.\ & - & $1.0$ & $1.0$ \\
opt.\ steps & 70 & 70 & 70 & 70\\
\multicolumn{5}{c}{}
\end{tabular}

\begin{tabular}{l|cccccc}
(b) size $n$ (qubits) & 5 & 7 & 9 & 11 & 13 & 23 \\
\hline
algorithms & all & all & all & all & all & HE-ITE \\
layers $p$ in HE-ITE & 1 & 1 & 1 & 1 & 1 & 1 \\
layers $p$ in all except HE-ITE & 2 & 3 & 4 & 5 & 6 & -\\
measurement shots $M$ & 10 & 50 & 100 & 150 & 200 & $2^{n_{\text{cone}} + 2}$\\
\multicolumn{7}{c}{}
\end{tabular}

\begin{tabular}{l|c}
\multicolumn{2}{l}{(c) 9-qubit experiment on Honeywell H1}\\
\hline
layers $p$ & 1 \\
ansatz circuit & Fig.~\ref{fig_exp}(b) \\
adaptive $\tau$ & yes, $g_c = 0.2$ \\
measurement shots $M$ & 500 \\
optimization steps & 9
\end{tabular}
\caption{\label{t_settings}
Simulation and experimental settings.
(a) The cost function for VQE and QAOA is the average energy for their respective ansatz circuits $\ket{\psi(\bm{\theta})}$ and $\ket{\psi(\bm{\gamma}, \bm{\beta})}$.
The learning rate for F-VQE is the inverse of the cost function's Hessian diagonal.
(b) For the 23-qubit problems HE-ITE uses $2^{n_{\text{cone}}+2}$ measurement shots for each causal cone, where $n_{\text{cone}}$ is the number of qubits in the causal cone.
(c) Different settings for the 9-qubit experiment.
In the ansatz circuit all rotation gates except the first and the last ones on each qubit are removed.
Additional details corresponding to the experiment are provided in Appendix~\ref{app_exp_det}.
}
\end{table}

\begin{figure}[t]
\centering
\includegraphics[width=65.344mm]{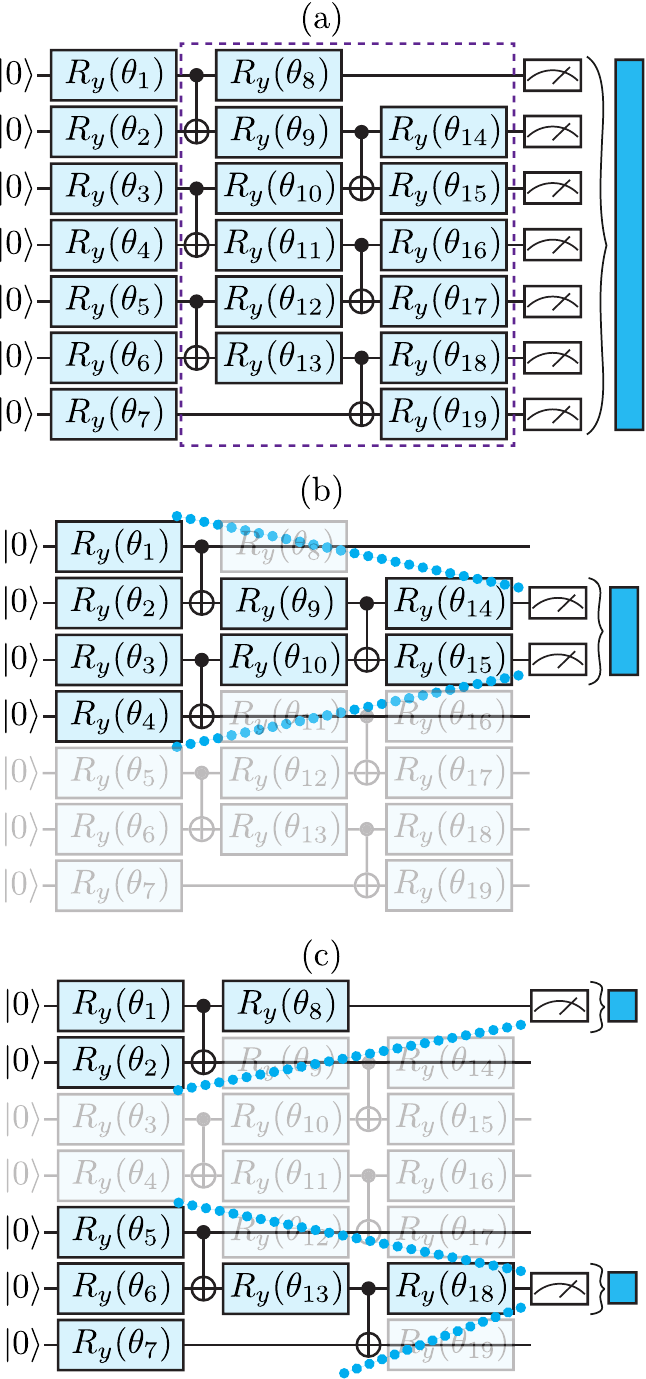}
\caption{\label{fig_ansatz}
Parameterized quantum circuit defined by a vector $\bm{\theta} = (\theta_1, \ldots, \theta_m)$ of $m$ parameters, here for $m = 19$.
(a) For a number $p$ of circuit layers the block of gates inside the dashed rectangle is repeated $p$ times.
Here we show the circuit for $p = 1$.
The F-VQE algorithm samples the entire circuit to evaluate a global observable indicated by the blue rectangle.
(b)-(c) Highlighted qubits and gates constitute the causal cone that HE-ITE uses to evaluate 2-local observables on two neighboring qubits and two non-neighboring qubits, respectively.
}
\end{figure}

\textit{F-VQE}.
This algorithm uses the parameterized quantum circuit shown in Fig.~\ref{fig_ansatz}(a).
An initial state $\ket{\psi_0} = \ket{+}^{\otimes n}$ is prepared by setting to $\pi/2$ the parameters in the last rotation on each qubit and setting the remaining parameters to $0$.
Parameters are iteratively updated using analytical gradient descent as described in Section~\ref{subsec:F-VQE}.
At each optimization step, the value of the parameter $\tau$ is adapted according to the procedure explained in Section~\ref{s_tau_sch}.
We choose a threshold of $g_c = 0.1$ for the gradient norm and solve the implicit equation with a precision $0 < g_c - g(\tau_t) < 0.01$.
For the learning rate $\eta$ we choose the inverse of the Hessian's diagonal.
As shown in Appendix~\ref{app_an_grad_fvqe}, at each optimization step $t$ all diagonal elements of the Hessian have the same value which can be computed by means of the circuit $\ket{\psi_{t-1}}$.
This is a quasi-Newton method~\cite{Dalgaard2020, Andrea2021} that uses only the diagonal of the Hessian matrix and can be realized without additional cost.

\textit{HE-ITE}.
In relation to using causal cones with F-VQE, we have chosen to concentrate our analysis on the exponential filter.
In this case the F-VQE method is equivalent to the HE-ITE algorithm called Angle Update in~\cite{BeFiLu20}.
We adapt this algorithm to the general QUBO Hamiltonians with long-range interactions considered here.
We use the ansatz circuit depicted in Fig.~\ref{fig_ansatz}(a) with $p = 1$ layer.
Figures~\ref{fig_ansatz}(b) and (c) show examples of causal cones in HE-ITE.
A total of $70$ time steps are performed with a fixed imaginary time step $\tau = 1.0$.
For the $23$-qubit problems, we choose the number of measurement shots dependent on the number $n_{\text{cone}}$ of qubits in the cone as $2^{n_{\text{cone}} + 2}$.

\begin{figure*}[t]
\centering
\includegraphics[width=\textwidth]{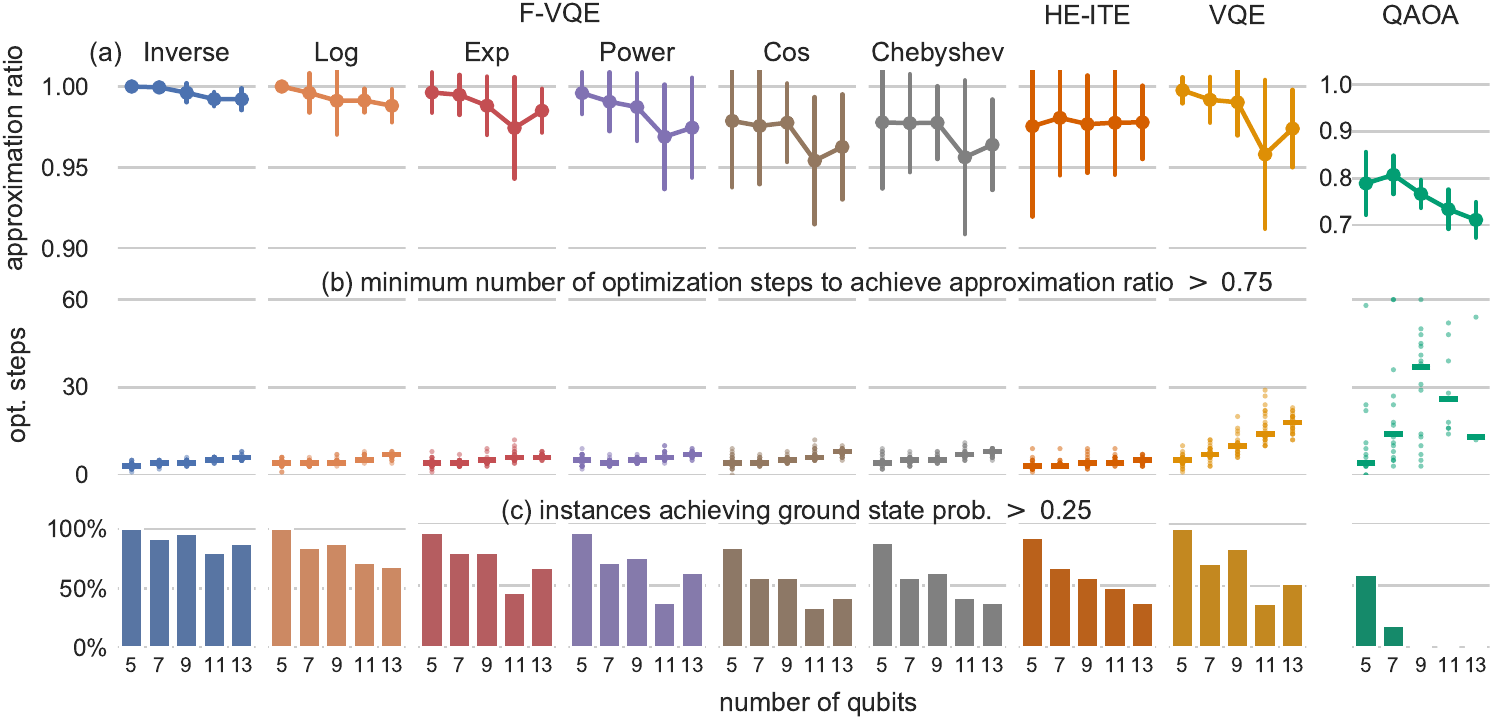}
\caption{\label{fig_app_rat}
Simulation results for 25 random weighted MaxCut problems of various sizes.
(a) Average approximation ratio (circles) and standard deviation (error bars).
Note the different range for QAOA.
(b) Optimization steps: median (lines) and all instances (dots).
The number of instances that achieve an approximation ratio of $0.75$ is 25 for all algorithms and problem sizes except for QAOA.
For QAOA these numbers are 17, 22, 21, 9, and 3 for $n = 5$, $7$, $9$, $11$, and $13$, respectively.
(c) Fraction of instances that achieve a probability of measuring the ground state above $0.25$.
}
\end{figure*}

\textit{VQE.}
For this algorithm we employ the same ansatz as in F-VQE and HE-ITE, shown in Fig.~\ref{fig_ansatz}(a).
The computation of the analytical gradient for this ansatz requires two quantum circuits per parameter as described in Appendix~\ref{app_an_grad_vqe}.
For the VQE cost function the diagonal of the Hessian can be obtained using the same circuits that are needed for the analytical gradient, similar to F-VQE.
Thus it is possible to apply the same quasi-Newton method to the VQE optimization.
However, we found that this heuristic performs worse than simply fixing a learning rate for all optimization steps and partial derivatives.
More specifically, in the simulations we found that the Hessian diagonal elements frequently vanish so that the parameter update often diverges with this heuristic.
The analysis of this phenomenon goes beyond the scope of this work and therefore we choose to fix a learning rate $\eta$ for all optimization steps and partial derivatives.
Comparing the performance of the values $\eta = 1$, $0.1$ and $0.01$ using our 5-qubit MaxCut instances, we conclude that $\eta = 1$ performs best and hence use this fixed value for $\eta$ in our simulations and experiments.

\textit{QAOA}.
The parameterized quantum circuit for QAOA is:
\begin{align}
    \ket{\psi(\bm{\gamma}, \bm{\beta})} &= U(\gamma_p, \beta_p) \cdots U(\gamma_1, \beta_1) \ket{+}^{\otimes n} ,\label{eq_qaoa_ansatz}\\
    U(\gamma_j, \beta_j) &= \exp \left( -i \gamma_j \sum_{q = 1}^n X_q \right) \exp \left( -i \beta_j \mathcal{H} \right) .\label{eq_qaoa_unitary}
\end{align}
Here $\ket{+} = (1/\sqrt{2})(\ket{0}+\ket{1})$, $X_{q}$ denotes the Pauli operator $X$ acting on qubit $q$, and the ansatz is defined by the $m = 2 p$ parameters in the vectors $\bm{\gamma} = (\gamma_1,\,\ldots,\,\gamma_p)$ and $\bm{\beta} = (\beta_1,\,\ldots,\,\beta_p)$.
We initialize the parameters randomly in the range $[0, \pi]$.
We optimize the parameters using analytical gradient descent.
As shown in Appendix~\ref{app_an_grad_qaoa}, the computation of the analytical gradient for the QAOA ansatz requires just the ansatz circuit with various parameter sets.
In the QAOA simulations we do not use the quasi-Newton method to determine the learning rate, as this would need additional circuits.
Instead we fix a learning rate $\eta$ for all optimization steps and partial derivatives.
We choose $\eta = 1$ as it is the best performing learning rate for the 5-qubit MaxCut instances where we compared the performance of $\eta = 1$, $0.1$, and $0.01$.

\subsection{Performance}
\label{subsec:performance}

In the following, we present and analyze the numerical and experimental results of the F-VQE algorithms and compare them with those of HE-ITE, VQE and QAOA.
For each MaxCut instance we pay special attention to two benchmark quantities: the approximation ratio and the probability of measuring the ground state.
These quantities have some dependence.
A large probability of sampling the ground state $P_\psi(\lambda_0) \approx 1$ implies a large approximation ratio $\langle \alpha \rangle_\psi \approx 1$.
However, a quantum state $\ket{\psi}$ given by a superposition of low-energy excited states can exhibit a large approximation ratio $\langle \alpha \rangle_\psi \approx 1$ but low ground state probability $P_\psi(\lambda_0) \approx 0$.

We compare the performance of various filtering operators in F-VQE in Fig.~\ref{fig_app_rat}.
Here filtering operators are sorted from left to right by performance and the inverse filter is the best performing one.
Figure~\ref{fig_app_rat}(a) shows for each algorithm considered in this article the final approximation ratio averaged over all 25 MaxCut instances for each problem size.
We observe that the best performing filters achieve the largest approximation ratios and are more reliable in obtaining such approximation ratios, which can be gathered from the small values of the corresponding standard deviations.
Figure~\ref{fig_app_rat}(b) shows the distribution of the minimum number of optimization steps required to achieve an approximation ratio above $0.75$.
Here all filters show a similar performance: they require 5 or fewer optimization steps with little deviation from the median.
Figure~\ref{fig_app_rat}(c) shows the fraction of MaxCut instances where the algorithms obtain a probability of sampling the ground state above $0.25$.
The probability of sampling it at least once with $M$ measurement shots is then $1 - (1 - 0.25)^M$.
The best filters achieve this probability for a larger fraction of MaxCut instances.

Let us now compare the best performing filtering operator, the inverse filter, with VQE and QAOA.
F-VQE requires less optimization steps than VQE and QAOA to achieve larger and more consistent approximation ratios.
This can be seen in Figs.~\ref{fig_app_rat}(a) and (b) as well as in Fig.~\ref{fig_main}(a).
Additionally, as shown in Fig.~\ref{fig_app_rat}(c), F-VQE converges to the ground state more often for all problem sizes.

\begin{figure}
\centering
\includegraphics[width=\linewidth]{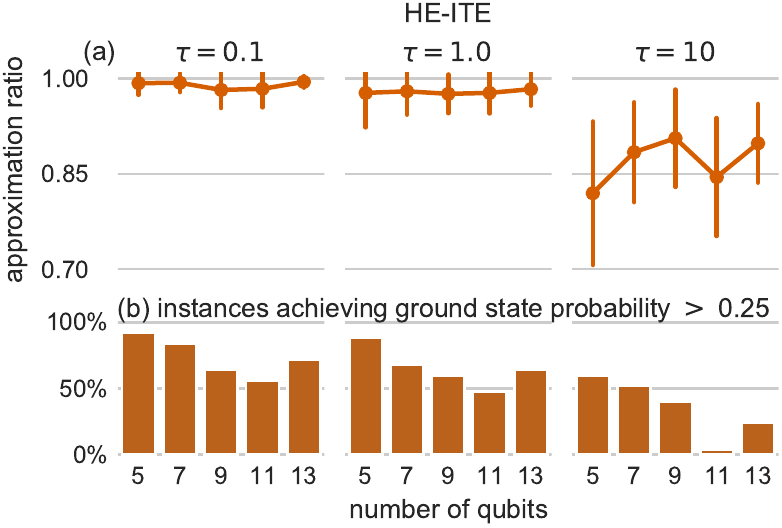}
\caption{\label{f_imaginary_time}
Simulation results for the HE-ITE algorithm applied to 25 random weighted MaxCut problems of various sizes using different imaginary time steps $\tau$ and total imaginary time $100$.
(a) Approximation ratio: average (circles) and standard deviation (error bars).
(b) Fraction of instances that achieve a probability of measuring the ground state above $0.25$.
}
\end{figure}

The HE-ITE algorithm also achieves approximation ratios close to the optimum and frequently converges to the ground state.
Moreover, the evolution of the approximation ratio shown in Fig.~\ref{fig_main}(a) almost overlaps with that for F-VQE.
Similarly Fig.~\ref{fig_main}(b) shows that HE-ITE obtains an average approximation ratio close to the optimum for 25 MaxCut instances of 23 qubits.
Importantly, the quantum circuits never use more than six qubits.
The inset in Fig.~\ref{fig_main}(b) shows the average fraction of circuits used for each qubit count.
We observe that the majority of circuits require just four qubits.

The performance of HE-ITE depends on the imaginary time step $\tau$ and to analyze it we have run HE-ITE for a long total imaginary time of $100$. Figure~\ref{f_imaginary_time} compares for three values of $\tau$ the final approximation ratios and the fraction of MaxCut instances where a probability of sampling the ground state above $0.25$ is achieved.
We conclude that HE-ITE improves systematically by choosing smaller values of $\tau$.

\begin{figure}
\centering
\includegraphics[width=0.9\linewidth]{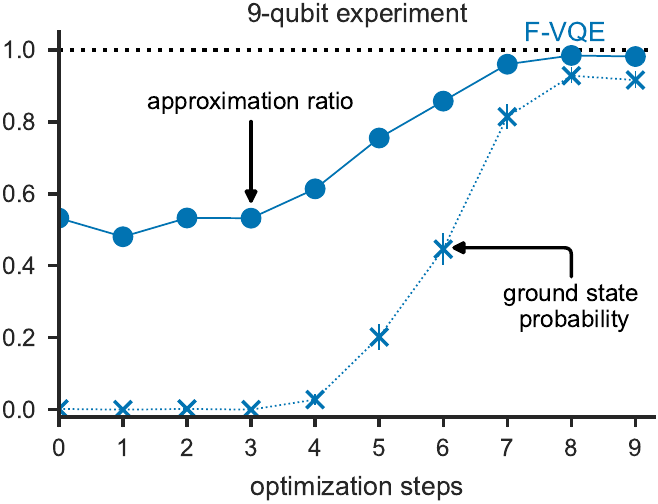}
\caption{\label{fig_honey}
Approximation ratio and probability of sampling the ground state for a single random weighted MaxCut instance solved on the Honeywell H1 trapped-ion quantum processor~\cite{Pino2021}. Error bars indicate one standard deviation. Details are provided in Appendix~\ref{app_exp_det}.
}
\end{figure}

To demonstrate the experimental feasibility, we run F-VQE with the inverse filter on the Honeywell H1 trapped-ion quantum processor and solve a random 9-qubit MaxCut instance.
Experimental settings (see Table~\ref{t_settings}(c)) are similar to those used for the numerical simulations, with some differences described in detail in Appendix~\ref{app_exp_det}.
The main difference is that we use only $p = 1$ layer and remove all rotation gates except the first and the last ones for each qubit.
Figure~\ref{fig_honey} shows the approximation ratio and the probability of measuring the ground state at each optimization step.
The final approximation ratio is $0.9844 \pm 0.0062$ and the probability of sampling the ground state after the final optimization step is $0.928 \pm 0.024$.
Here the value after $\pm$ indicates a $95\%$ confidence interval.

\section{Conclusions and outlook}
\label{s_conc_out}

We have introduced variational quantum algorithms that make use of filtering operators to solve combinatorial optimization problems.
The Quantum Variational Filtering (QVF) algorithm uses a parameterized quantum circuit to approximate the repeated action of a filtering operator.
Filtering VQE (F-VQE) is a particularly efficient version of QVF and similar to VQE.
These algorithms impose no restrictions on the ansatz circuit so that we can choose the one that performs optimally on hardware.
We have also tested a hardware-efficient imaginary time evolution (HE-ITE) algorithm introduced in~\cite{BeFiLu20}.
This algorithm approximates the action of the imaginary time evolution operator.
Using causal cones, HE-ITE can require circuits that are significantly smaller than the problem size.

We have compared F-VQE and HE-ITE with VQE and QAOA using a set of random weighted MaxCut problems of various sizes.
The F-VQE algorithm achieved larger and more consistent approximation ratios as well as more reliable convergence to the optimal solutions than VQE and QAOA via fewer optimization steps.
The HE-ITE algorithm obtained similar approximation ratios and ground state convergence with drastically reduced qubit count.
Moreover, F-VQE successfully solved a 9-qubit MaxCut problem on the Honeywell H1 trapped-ion quantum processor~\cite{Pino2021}.
We conclude that F-VQE and HE-ITE are powerful algorithms to solve combinatorial optimization problems on noisy quantum computers.

Owing to the high flexibility of F-VQE, various promising strategies can be considered to further improve the performance.
F-VQE can readily be combined with the Conditional Value-at-Risk cost function \cite{Barkoutsos2020, Kolotouros2021, Diez2021} to provide new filtering operators with additional capabilities.
Local cost functions and shallow ansatz circuits can be used to avoid barren plateaus~\cite{Cerezo2021}.
F-VQE might also benefit from the original QAOA ansatz~\cite{Farhi2014} or generalizations thereof such as the Hamiltonian variational ansatz~\cite{WeHaTr15, Wiersema2020}, the quantum alternating operator ansatz~\cite{HaEtAl19}, the hardware-efficient mixer-phaser ansatz~\cite{LaRose2021} or the depth optimized QAOA ansatz~\cite{MaEtAl21}.
The ansatz can be specifically selected to minimize experimental noise on quantum hardware~\cite{Du2020}.
To this end, the quantum autoencoder is a powerful concept~\cite{RoOlAs17, ChWa21}.
It is enticing to combine F-VQE with the holographic ansatz~\cite{FoEtAl20} that has led to impressive results on the Honeywell quantum computer~\cite{FoEtAl21, ChEtAl21}.
Any ansatz can be enhanced by extending it with classical neural networks~\cite{ZhEtAl21}.
The classical optimizer for the parameters can simultaneously adjust the ansatz circuit structure to alleviate the effects of experimental noise~\cite{Ostaszewski_Benedetti_2021, Ostaszewski2021}.
It might also be beneficial to grow the ansatz during the optimization as in ADAPT-VQE~\cite{GrEtAl19}, Adapt-QAOA~\cite{ZhEtAl20}, layerwise learning~\cite{SkEtAl21} or Layer VQE~\cite{Liu2021}.
Advanced gradient-descent techniques like stochastic gradient descent may reduce the number of optimization steps and avoid local minima~\cite{Sweke2020}.
Finally, different heuristic adaptations of the parameter $\tau$ and the learning rate can be explored to gain more information from circuit samples.

An interesting future application for QVF and F-VQE is the computation of states different from the ground state.
Our filtering algorithms can target any state of energy $E_{\text{target}}$, e.g.\ a specific excited state, simply by using the slightly modified filtering operator $f(|\mathcal{H}-E_{\text{target}}|; \tau)$.

Another interesting application for QVF and F-VQE is the optimization based on black-box cost functions.
Our filtering algorithms (as well as VQE) can optimize a variational ansatz given any black-box cost function $\mathcal{H}_{\text{black-box}}(\bm{x})$ that takes as input a bitstring of binary variables $\bm{x}$ and returns the associated cost function value.
Therefore it is not necessary to represent the problem in terms of a QUBO Hamiltonian.
A different problem representation might, e.g., reduce the required qubit count.
We note that one of the first successful approaches for the optimization corresponding to such black-box cost functions is based on simulated annealing~\cite{KiGeVe83} which made it possible to tackle real-world problems that could not be tackled before~\cite{Ro85}.
An interesting future project is the comparison of our filtering algorithms with simulated annealing.

It is exciting to think about using filtering operators for the computation of ground states of quantum Hamiltonians.
This can be achieved with some filtering operators.
For example, the imaginary time evolution filter is often used in combination with quantum Hamiltonians, for details see~\cite{BeFiLu20} and references therein.
Using the circuits described in this article, it is also straightforward to realize the power and Chebyshev filters for integer values of $\tau$ in QVF as well as F-VQE.
With respect to the inverse filter -- the best performing filtering operator of this work -- there exists a proposal to realize it with quantum Hamiltonians by means of a Fourier approximation~\cite{Kyriienko2020}.
With the help of the Fourier transform, also other filtering operators can be applied to quantum Hamiltonians~\cite{ZeSuYu21}.

The insights gained here are also useful for the design of new quantum-inspired classical algorithms that can obtain significant speedups compared to traditional classical methods~\cite{AlPe21, Ga21, Patti2021}.
Here one interesting question is whether tensor network methods for ground state minimization can benefit from the use of filtering operators.
Faster and more accurate ground state algorithms are crucial, e.g., for the construction of systematically improved functionals for density functional theory~\cite{LuEtAl16}, or to further improve fast solvers for the boundary value problem of general nonlinear partial differential equations~\cite{LuMoJa18}.
Such algorithms can also give us new answers to traditional tensor network questions regarding quantum phases and phase transitions in strongly correlated quantum systems~\cite{Or14, CiEtAl20}.

\section{Acknowledgments}
We thank Brian Neyenhuis and all the Honeywell Quantum Solutions team for their availability and support with the H1 device.
We acknowledge the cloud computing resources received from the `Microsoft for Startups’ program. We also thank Seyon Sivarajah, Alec Edginton, Richie Yeung, Ross Duncan, and all the TKET development team for their technical support.

\appendix

\section{Hadamard test for QVF}
\label{app_hadamard}

In this Appendix we describe how to minimize the QVF cost function in Eq.~\eqref{eq_cost_fun} with the help of a Hadamard test.

\begin{figure*}
\centering
\includegraphics[width=134.901mm]{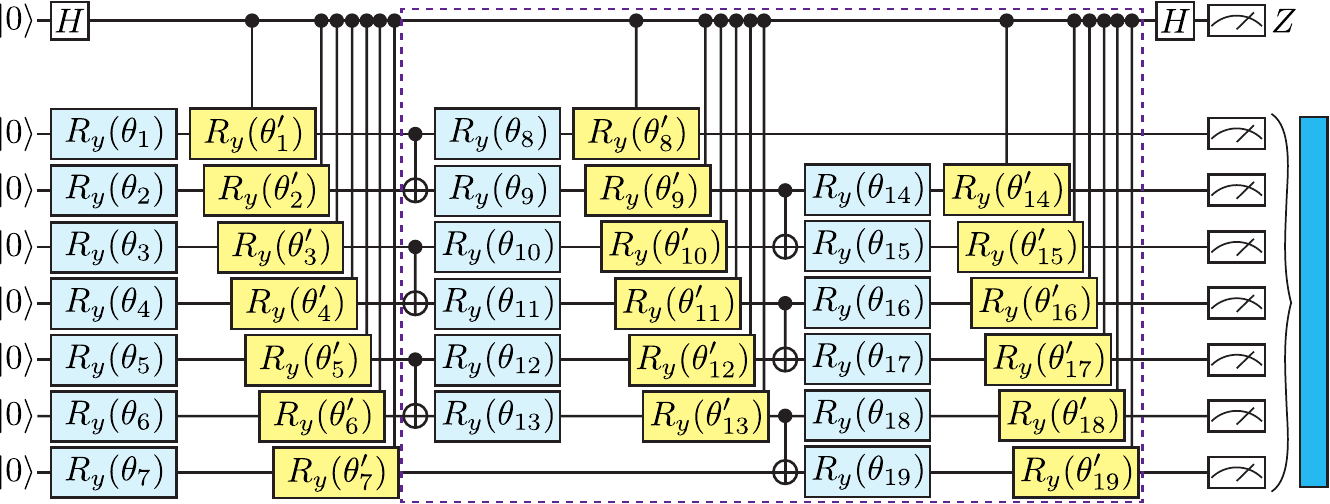}
\caption{\label{fig_hadamard}
Hadamard test circuit to compute $\Re \bra{\psi(\bm{\phi})} F_t \ket{\psi(\bm{\theta})}$ for the parameterized quantum circuit in Fig.~\ref{fig_ansatz}(a).
The $H$ gates acting on the ancilla qubit at the top are Hadamard gates.
For an ansatz of $p$ layers the block of gates inside the dashed rectangle is applied $p$ times.
This figure shows $p = 1$ layer.
The parameter vectors satisfy $\bm{\theta}' = \bm{\phi} - \bm{\theta}$.
}
\end{figure*}

Evaluating the cost function in Eq.~\eqref{eq_cost_fun} requires computing a quantity of the form $\bra{\psi(\bm{\phi})} F_t \ket{\psi(\bm{\theta})}$, where $\ket{\psi(\bm{\theta})}$ and $\ket{\psi(\bm{\phi})}$ are $n$-qubit quantum ansatz circuits shown in Fig.~\ref{fig_ansatz}(a) with parameter vectors $\bm{\theta}$ and $\bm{\phi}$.
The computation of such a quantity is also needed for the gradient calculation in Eq.~\eqref{eq_grad}.
This quantity can be computed with a Hadamard test by evaluating the expectation value of the diagonal and Hermitian observable $Z_\text{anc} \otimes F_t$ with the specific circuit $W(\bm{\theta}, \bm{\phi})$ in Fig.~\ref{fig_hadamard} -- composed of one ancilla qubit labeled by $\text{anc}$ and a $n$-qubit register -- that implements the following operation:
\begin{eqnarray}
W(\bm{\theta}, \bm{\phi}) \ket{0}_\text{anc} \ket{\bm{0}} & = & \frac{1}{2} \ket{0}_\text{anc} ( \ket{\psi(\bm{\theta})} + \ket{\psi(\bm{\phi})} ) +\nonumber\\
& & \frac{1}{2} \ket{1}_\text{anc} ( \ket{\psi(\bm{\theta})} - \ket{\psi(\bm{\phi})} ) .
\end{eqnarray}

\section{Analytical derivatives}
\label{app_an_grad}

In this Appendix we derive the analytical gradient in Eq.~\eqref{eq_grad} and the diagonal elements of the Hessian matrix for the cost function in Eq.~\eqref{eq_cost_fun} corresponding to the QVF algorithm.
We use the parameter-shift rule~\cite{Mitarai2018, Schuld2019} to derive the analytical gradient in Eq.~\eqref{eq_grad_simp} for the F-VQE algorithm.
This procedure is also used to derive analytical gradients for VQE and QAOA.

\subsection{QVF and F-VQE}
\label{app_an_grad_fvqe}

In the quantum ansatz circuits $\ket{\psi(\bm{\theta})}$ of Fig.~\ref{fig_ansatz}(a) parameters $\bm{\theta} = (\theta_1,\,\theta_2,\ldots,\theta_m)$ are present only in rotation gates of the form $R_G(\theta_j) = \exp(-i \theta_j G/2)$, where $G$ is a single-qubit Pauli operator or a tensor product of Pauli operators.
As a function of a single parameter, the circuit can be expressed in terms of two fixed unitaries $V_A,\,V_B$ and the rotation gate corresponding to the parameter:
\begin{equation}
\ket{\psi(\theta_j)} = V_A R_G(\theta_j) V_B \ket{\bm{0}} ,
\end{equation}
where $\ket{\bm{0}} = \ket{0}^{\otimes n}$ is the initial register state.
Hence the first and second derivatives with respect to a parameter $\theta_j$ are:
\begin{align}
\dfrac{\partial \ket{\psi(\theta_j)}}{\partial \theta_j} &= \frac{1}{2} V_A R_G(\theta_j)(-i G) V_B \ket{\bm{0}} = \frac{1}{2} \ket{\psi(\theta_j + \pi)}\\
\dfrac{\partial^2 \ket{\psi(\theta_j)}}{\partial \theta_j^2} &= \frac{1}{4} V_A R_G(\theta_j) (-i G)^2 V_B \ket{\bm{0}} = -\frac{1}{4} \ket{\psi(\theta_j)} ,
\end{align}
where we have used that $-i G = R_G(\pi)$.

Then the first and second derivatives of the cost function in Eq.~\eqref{eq_cost_fun} are:
\begin{align}
\dfrac{\partial \mathcal{C}_t(\bm{\theta})}{\partial \theta_j} &= -\frac{\Re \bra{\psi_{t-1}} F_t \ket{\psi(\bm{\theta} + \pi \bm{e}_j)}}{2 \sqrt{\langle F_t^2 \rangle_{\psi_{t-1}}}} ,\\ \label{second_der_fvqe}
\dfrac{\partial^2 \mathcal{C}_t(\bm{\theta})}{\partial \theta_j^2} &= \frac{\Re \bra{\psi_{t-1}} F_t \ket{\psi(\bm{\theta})}}{4 \sqrt{\langle F_t^2 \rangle_{\psi_{t-1}}}} .
\end{align}
The first equation corresponds to Eq.~\eqref{eq_grad}.
Since the filtering operator $F_t$ is Hermitian, when the first equation is evaluated for the vector of parameters $\bm{\theta}_{t-1}$ that produces the state $\ket{\psi_{t-1}} = \ket{\psi(\bm{\theta}_{t-1})}$ we can use the parameter-shift rule to express the numerator as a sum of two circuits, which leads to Eq.~\eqref{eq_grad_simp}.
When the second equation is evaluated for $\bm{\theta}_{t-1}$, it results in:
\begin{align}  \label{eq_second_der}
\left.\dfrac{\partial^2 \mathcal{C}_t(\bm{\theta})}{\partial \theta_j^2}\right|_{\bm{\theta}_{t-1}} &= \frac{\langle F_t \rangle_{\psi_{t-1}}}{4 \sqrt{\langle F_t^2 \rangle_{\psi_{t-1}}}} ,
\end{align}
which requires only the quantum circuit for $\ket{\psi_{t-1}}$.

\subsection{VQE}
\label{app_an_grad_vqe}

The cost function is the average energy $\langle \mathcal{H} \rangle_{\psi(\bm{\theta})}$ and the ansatz circuit $\ket{\psi(\bm{\theta})}$ is the same as the one employed by F-VQE shown in Fig.~\ref{fig_ansatz}(a).
Given that each parameter is present in only one rotation gate, the parameter-shift rule can be applied to express the analytical gradient as:
\begin{equation}
\label{eq_vqe_an_grad}
\left.\dfrac{\partial \langle \mathcal{H} \rangle_{\psi(\bm{\theta})}}{\partial \theta_j}\right|_{\bm{\theta}_{t-1}} = \frac{1}{2} \left(\langle \mathcal{H} \rangle_{\psi_{t-1}^{j+}} - \langle \mathcal{H} \rangle_{\psi_{t-1}^{j-}}\right) .
\end{equation}
As with F-VQE, the circuits $\ket{\psi_{t-1}^{j\pm}} \equiv \ket{\psi\left(\bm{\theta}_{t-1} \pm \frac{\pi}{2}\bm{e}_j\right)}$ are implemented by shifting the parameter $\theta_j$ by an amount $\pm \pi/2$.

Using the same methods, the second derivative with respect to each parameter can be evaluated as
\begin{equation}
\label{eq_second_der_vqe}
\left.\dfrac{\partial^2 \langle \mathcal{H} \rangle_{\psi(\bm{\theta})}}{\partial \theta_j^2}\right|_{\bm{\theta}_{t-1}} = \frac{1}{2} \langle \mathcal{H} \rangle_{\psi_{t-1}^{j++}} - \frac{1}{2} \langle \mathcal{H} \rangle_{\psi_{t-1}},
\end{equation}
where $\ket{\psi_{t-1}} = \ket{\psi(\bm{\theta}_{t-1})}$ is the state with no shifts, and $\ket{\psi_{t-1}^{j++}} = \ket{\psi(\bm{\theta}_{t-1} + \pi\bm{e}_j)}$ is the state with a $+\pi$ shift.

\subsection{QAOA}
\label{app_an_grad_qaoa}

The QAOA ansatz in Eq.~\eqref{eq_qaoa_ansatz} is not of the previous form: here parameters multiply sums of tensor products of Pauli operators so that the partial derivatives become sums of circuit evaluations.
Given a Hamiltonian $\mathcal{H} = \sum_{k = 1}^K h_k Z_{Q_k}$ with real coefficients $h_k$ and $Z_{Q_k} = \bigotimes_{q \in Q_k} Z_q$, the ansatz derivatives are:
\begin{align}
\dfrac{\partial \ket{\psi(\bm{\gamma}, \bm{\beta})}}{\partial \gamma_j} &= \sum_{q = 1}^n \tilde{U}_{(j, p)}  (-i X_q) \tilde{U}_{(1, j-1)} \ket{+}^{\otimes n} ,\\
\dfrac{\partial \ket{\psi(\bm{\gamma}, \bm{\beta})}}{\partial \beta_j} &= \sum_{k = 1}^K h_k \tilde{U}_{(j+1, p)} (-i Z_{Q_k}) \tilde{U}_{(1, j)} \ket{+}^{\otimes n} .
\end{align}
Here $\tilde{U}_{a, b} = U(\bm{\gamma}_b, \bm{\beta}_b) \cdots U(\bm{\gamma}_{a+1}, \bm{\beta}_{a+1}) U(\bm{\gamma}_a, \bm{\beta}_a)$, where $U(\bm{\gamma}_j, \bm{\beta}_j)$ is defined in Eq.~\eqref{eq_qaoa_unitary}, contains all QAOA circuit layers from $a$ to $b$.
Therefore the partial derivatives of the QAOA cost function are:
\begin{align}
\dfrac{\partial \langle \mathcal{H} \rangle_{\psi(\bm{\gamma}, \bm{\beta})}}{\partial \gamma_j} &= 2\Re \bra{\psi(\bm{\gamma}, \bm{\beta})} \mathcal{H} \dfrac{\partial \ket{\psi(\bm{\gamma}, \bm{\beta})}}{\partial \gamma_j} ,\\
\dfrac{\partial \langle \mathcal{H} \rangle_{\psi(\bm{\gamma}, \bm{\beta})}}{\partial \beta_j} &= 2\Re \bra{\psi(\bm{\gamma}, \bm{\beta})} \mathcal{H} \dfrac{\partial \ket{\psi(\bm{\gamma}, \bm{\beta})}}{\partial \beta_j} .
\end{align}
If the parameter-shift rule is applied to every term individually, we obtain the analytical gradient for the QAOA ansatz in terms of ansatz circuits for various parameter sets:
\begin{align}
\dfrac{\partial \langle \mathcal{H} \rangle_{\psi(\bm{\gamma}, \bm{\beta})}}{\partial \gamma_j} &= \sum_{q = 1}^n \left( \langle \mathcal{H} \rangle_{\psi_X^{(j, q)+}} - \langle \mathcal{H} \rangle_{\psi_X^{(j, q)-}} \right) ,\label{eq_qaoa_grad_gamma}\\
\dfrac{\partial \langle \mathcal{H} \rangle_{\psi(\bm{\gamma}, \bm{\beta})}}{\partial \beta_j} &= \sum_{k = 1}^K h_k \left( \langle \mathcal{H} \rangle_{\psi_{\mathcal{H}}^{(j, k)+}} - \langle \mathcal{H} \rangle_{\psi_{\mathcal{H}}^{(j, k)-}} \right) .\label{eq_qaoa_grad_beta}
\end{align}
The evaluation of all gradient components requires the following $2 p (n + K)$ circuits that are defined by inserting a rotation gate in the ansatz circuit:
\begin{align}
\ket{\psi_X^{(j, q)\pm}} &= \tilde{U}_{(j, p)} R_{X_q} \left( \pm \frac{\pi}{2} \right) \tilde{U}_{(1, j-1)} \ket{+}^{\otimes n} ,\\
\ket{\psi_{\mathcal{H}}^{(j, k)\pm}} &= \tilde{U}_{(j+1, p)} R_{Z_{Q_k}} \left( \pm \frac{\pi}{2} \right) \tilde{U}_{(1, j)} \ket{+}^{\otimes n} .
\end{align}

\section{Experimental details}
\label{app_exp_det}

This Appendix provides further details on the 9-qubit experimental results, shown in Fig.~\ref{fig_honey}, obtained with the Honeywell H1 trapped-ion quantum processor solving a MaxCut problem.

\begin{figure*}[t]
\centering
\includegraphics[width=0.8\textwidth]{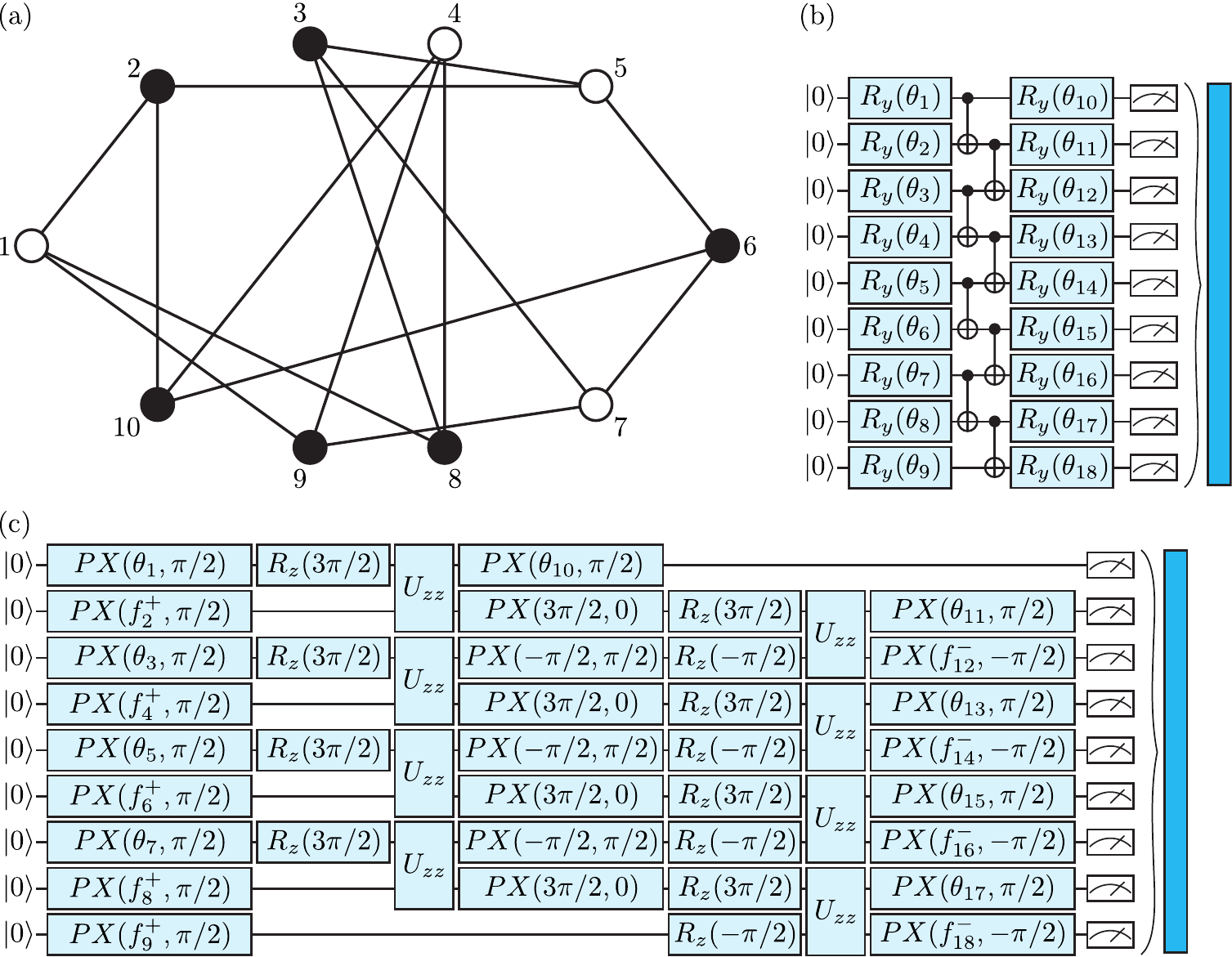}
\caption{\label{fig_exp}
Experimental details relating to Fig.~\ref{fig_honey}. (a) 10-node 3-regular weighted graph for the MaxCut problem with weights given in Tab.~\ref{t_weights}.
The corresponding solution is the division into black and white nodes.
(b) Parameterized quantum circuit employed by F-VQE and (c) its compiled counterpart expressed in terms of Honeywell H1 native gates.
}
\end{figure*}

The considered MaxCut problem is defined on the 10-node 3-regular weighted graph depicted in Fig.~\ref{fig_exp}(a).
The corresponding MaxCut weights are given in Tab.~\ref{t_weights}.
The solution divides the nodes into the two sets of black and white nodes shown in Fig.~\ref{fig_exp}(a), equivalent to the 10-variable solution vector $(1, -1, -1, 1, 1, -1, 1, -1, -1, -1)$ which corresponds to the 9-qubit solution computational state $\ket{011001011}$ in our experiment, as explained in Section~\ref{subsec:problem_hamiltonian}.

Figure~\ref{fig_exp}(b) shows the parameterized quantum circuit that is used in the experiment.
This circuit is simpler than the ones that are used for the simulations -- cf.~Fig.~\ref{fig_ansatz}(a) -- as it contains just $p = 1$ layer with two single-qubit rotation gates per qubit, one at the beginning and one at end of the circuit.

Before the circuit of Fig.~\ref{fig_exp}(b) is run on the Honeywell H1 processor, it is compiled into the circuit shown in Fig.~\ref{fig_exp}(c) which is composed of the native gates of the processor.
There are three native gates~\cite{Pino2021}: the standard single-qubit rotation gate around the $Z$ axis, a single-qubit rotation gate around an axis in the $X$-$Y$ plane $PX(\theta, \phi) = R_{z}(\phi) R_{x}(\theta) R_{z}(-\phi)$, and the two-qubit interaction gate $U_{zz} = \exp(-i \pi Z \otimes Z / 4)$.
The compilation of the original two-qubit gates in  Fig.~\ref{fig_exp}(b) changes the rotation angles of the single-qubit gates to
\begin{equation}
 f^{\pm}_{j} = 2 \text{atan2} \left( -\cos{\frac{\theta_{j}}{2}} \pm \sin{\frac{\theta_{j}}{2}}, \cos{\frac{\theta_{j}}{2}} \pm \sin{\frac{\theta_{j}}{2}} \right) .
\end{equation}

\begin{table}
\begin{tabular}{c|c}
edge & weight\\
\hline
(1, 2) & 0.0609\\
(1, 8) & 0.1574\\
(1, 9) & 0.1392\\
(2, 5) & 0.6131\\
(2, 10) & 0.2670
\end{tabular}
\hspace{4mm}
\begin{tabular}{c|c}
edge & weight\\
\hline
(3, 5) & 0.4156\\
(3, 7) & 0.2020\\
(3, 8) & 0.0120\\
(4, 8) & 0.6738\\
(4, 9) & 0.5296
\end{tabular}
\hspace{4mm}
\begin{tabular}{c|c}
edge & weight\\
\hline
(4, 10) & 0.9927\\
(5, 6) & 0.2617\\
(6, 7) & 0.7781\\
(6, 10) & 0.0202\\
(7, 9) & 0.3973
\end{tabular}
\caption{\label{t_weights}
MaxCut weights for the MaxCut problem considered in our experiment based on the graph in Fig.~\ref{fig_exp}(a).
}
\end{table}

\bibliography{bibliography}

\begin{thebibliography}{76}%
\makeatletter
\providecommand \@ifxundefined [1]{%
 \@ifx{#1\undefined}
}%
\providecommand \@ifnum [1]{%
 \ifnum #1\expandafter \@firstoftwo
 \else \expandafter \@secondoftwo
 \fi
}%
\providecommand \@ifx [1]{%
 \ifx #1\expandafter \@firstoftwo
 \else \expandafter \@secondoftwo
 \fi
}%
\providecommand \natexlab [1]{#1}%
\providecommand \enquote  [1]{``#1''}%
\providecommand \bibnamefont  [1]{#1}%
\providecommand \bibfnamefont [1]{#1}%
\providecommand \citenamefont [1]{#1}%
\providecommand \href@noop [0]{\@secondoftwo}%
\providecommand \href [0]{\begingroup \@sanitize@url \@href}%
\providecommand \@href[1]{\@@startlink{#1}\@@href}%
\providecommand \@@href[1]{\endgroup#1\@@endlink}%
\providecommand \@sanitize@url [0]{\catcode `\\12\catcode `\$12\catcode
  `\&12\catcode `\#12\catcode `\^12\catcode `\_12\catcode `\%12\relax}%
\providecommand \@@startlink[1]{}%
\providecommand \@@endlink[0]{}%
\providecommand \url  [0]{\begingroup\@sanitize@url \@url }%
\providecommand \@url [1]{\endgroup\@href {#1}{\urlprefix }}%
\providecommand \urlprefix  [0]{URL }%
\providecommand \Eprint [0]{\href }%
\providecommand \doibase [0]{https://doi.org/}%
\providecommand \selectlanguage [0]{\@gobble}%
\providecommand \bibinfo  [0]{\@secondoftwo}%
\providecommand \bibfield  [0]{\@secondoftwo}%
\providecommand \translation [1]{[#1]}%
\providecommand \BibitemOpen [0]{}%
\providecommand \bibitemStop [0]{}%
\providecommand \bibitemNoStop [0]{.\EOS\space}%
\providecommand \EOS [0]{\spacefactor3000\relax}%
\providecommand \BibitemShut  [1]{\csname bibitem#1\endcsname}%
\let\auto@bib@innerbib\@empty
\bibitem [{\citenamefont {Korte}\ and\ \citenamefont {Vygen}(2018)}]{KoVy06}%
  \BibitemOpen
  \bibfield  {author} {\bibinfo {author} {\bibfnamefont {B.}~\bibnamefont
  {Korte}}\ and\ \bibinfo {author} {\bibfnamefont {J.}~\bibnamefont {Vygen}},\
  }\href {https://dl.acm.org/doi/book/10.5555/3241990} {\emph {\bibinfo {title}
  {Combinatorial Optimization: Theory and Algorithms}}},\ \bibinfo {edition}
  {6th}\ ed.\ (\bibinfo  {publisher} {Springer Publishing Company,
  Incorporated},\ \bibinfo {year} {2018})\BibitemShut {NoStop}%
\bibitem [{\citenamefont {Benedetti}\ \emph {et~al.}(2019)\citenamefont
  {Benedetti}, \citenamefont {Lloyd}, \citenamefont {Sack},\ and\ \citenamefont
  {Fiorentini}}]{Benedetti_2019}%
  \BibitemOpen
  \bibfield  {author} {\bibinfo {author} {\bibfnamefont {M.}~\bibnamefont
  {Benedetti}}, \bibinfo {author} {\bibfnamefont {E.}~\bibnamefont {Lloyd}},
  \bibinfo {author} {\bibfnamefont {S.}~\bibnamefont {Sack}},\ and\ \bibinfo
  {author} {\bibfnamefont {M.}~\bibnamefont {Fiorentini}},\ }\bibfield  {title}
  {\bibinfo {title} {Parameterized quantum circuits as machine learning
  models},\ }\href {https://doi.org/10.1088/2058-9565/ab4eb5} {\bibfield
  {journal} {\bibinfo  {journal} {Quantum Sci. Technol.}\ }\textbf {\bibinfo
  {volume} {4}},\ \bibinfo {pages} {043001} (\bibinfo {year}
  {2019})}\BibitemShut {NoStop}%
\bibitem [{\citenamefont {Cerezo}\ \emph
  {et~al.}(2021{\natexlab{a}})\citenamefont {Cerezo}, \citenamefont
  {Arrasmith}, \citenamefont {Babbush}, \citenamefont {Benjamin}, \citenamefont
  {Endo}, \citenamefont {Fujii}, \citenamefont {McClean}, \citenamefont
  {Mitarai}, \citenamefont {Yuan}, \citenamefont {Cincio},\ and\ \citenamefont
  {Coles}}]{CeEtAl20}%
  \BibitemOpen
  \bibfield  {author} {\bibinfo {author} {\bibfnamefont {M.}~\bibnamefont
  {Cerezo}}, \bibinfo {author} {\bibfnamefont {A.}~\bibnamefont {Arrasmith}},
  \bibinfo {author} {\bibfnamefont {R.}~\bibnamefont {Babbush}}, \bibinfo
  {author} {\bibfnamefont {S.~C.}\ \bibnamefont {Benjamin}}, \bibinfo {author}
  {\bibfnamefont {S.}~\bibnamefont {Endo}}, \bibinfo {author} {\bibfnamefont
  {K.}~\bibnamefont {Fujii}}, \bibinfo {author} {\bibfnamefont {J.~R.}\
  \bibnamefont {McClean}}, \bibinfo {author} {\bibfnamefont {K.}~\bibnamefont
  {Mitarai}}, \bibinfo {author} {\bibfnamefont {X.}~\bibnamefont {Yuan}},
  \bibinfo {author} {\bibfnamefont {L.}~\bibnamefont {Cincio}},\ and\ \bibinfo
  {author} {\bibfnamefont {P.~J.}\ \bibnamefont {Coles}},\ }\bibfield  {title}
  {\bibinfo {title} {Variational quantum algorithms},\ }\href
  {https://doi.org/10.1038/s42254-021-00348-9} {\bibfield  {journal} {\bibinfo
  {journal} {Nat. Rev. Phys.}\ }\textbf {\bibinfo {volume} {3}},\ \bibinfo
  {pages} {625} (\bibinfo {year} {2021}{\natexlab{a}})}\BibitemShut {NoStop}%
\bibitem [{\citenamefont {Bravo-Prieto}\ \emph
  {et~al.}(2020{\natexlab{a}})\citenamefont {Bravo-Prieto}, \citenamefont
  {Garc\'{\i}a-Mart\'{\i}n},\ and\ \citenamefont {Latorre}}]{Prieto2020}%
  \BibitemOpen
  \bibfield  {author} {\bibinfo {author} {\bibfnamefont {C.}~\bibnamefont
  {Bravo-Prieto}}, \bibinfo {author} {\bibfnamefont {D.}~\bibnamefont
  {Garc\'{\i}a-Mart\'{\i}n}},\ and\ \bibinfo {author} {\bibfnamefont {J.~I.}\
  \bibnamefont {Latorre}},\ }\bibfield  {title} {\bibinfo {title} {Quantum
  singular value decomposer},\ }\href
  {https://doi.org/10.1103/PhysRevA.101.062310} {\bibfield  {journal} {\bibinfo
   {journal} {Phys. Rev. A}\ }\textbf {\bibinfo {volume} {101}},\ \bibinfo
  {pages} {062310} (\bibinfo {year} {2020}{\natexlab{a}})}\BibitemShut
  {NoStop}%
\bibitem [{\citenamefont {Bharti}\ \emph {et~al.}(2021)\citenamefont {Bharti},
  \citenamefont {Cervera-Lierta}, \citenamefont {Kyaw}, \citenamefont {Haug},
  \citenamefont {Alperin-Lea}, \citenamefont {Anand}, \citenamefont {Degroote},
  \citenamefont {Heimonen}, \citenamefont {Kottmann}, \citenamefont {Menke},
  \citenamefont {Mok}, \citenamefont {Sim}, \citenamefont {Kwek},\ and\
  \citenamefont {Aspuru-Guzik}}]{BhEtAl21}%
  \BibitemOpen
  \bibfield  {author} {\bibinfo {author} {\bibfnamefont {K.}~\bibnamefont
  {Bharti}}, \bibinfo {author} {\bibfnamefont {A.}~\bibnamefont
  {Cervera-Lierta}}, \bibinfo {author} {\bibfnamefont {T.~H.}\ \bibnamefont
  {Kyaw}}, \bibinfo {author} {\bibfnamefont {T.}~\bibnamefont {Haug}}, \bibinfo
  {author} {\bibfnamefont {S.}~\bibnamefont {Alperin-Lea}}, \bibinfo {author}
  {\bibfnamefont {A.}~\bibnamefont {Anand}}, \bibinfo {author} {\bibfnamefont
  {M.}~\bibnamefont {Degroote}}, \bibinfo {author} {\bibfnamefont
  {H.}~\bibnamefont {Heimonen}}, \bibinfo {author} {\bibfnamefont {J.~S.}\
  \bibnamefont {Kottmann}}, \bibinfo {author} {\bibfnamefont {T.}~\bibnamefont
  {Menke}}, \bibinfo {author} {\bibfnamefont {W.-K.}\ \bibnamefont {Mok}},
  \bibinfo {author} {\bibfnamefont {S.}~\bibnamefont {Sim}}, \bibinfo {author}
  {\bibfnamefont {L.-C.}\ \bibnamefont {Kwek}},\ and\ \bibinfo {author}
  {\bibfnamefont {A.}~\bibnamefont {Aspuru-Guzik}},\ }\href@noop {} {\bibinfo
  {title} {{Noisy Intermediate-Scale Quantum (NISQ) algorithms}}} (\bibinfo
  {year} {2021}),\ \Eprint {https://arxiv.org/abs/2101.08448} {arXiv:2101.08448
  [quant-ph]} \BibitemShut {NoStop}%
\bibitem [{\citenamefont {Fern{\'{a}}ndez-Lorenzo}\ \emph
  {et~al.}(2021)\citenamefont {Fern{\'{a}}ndez-Lorenzo}, \citenamefont
  {Porras},\ and\ \citenamefont {Garc{\'{\i}}a-Ripoll}}]{Lorenzo2021}%
  \BibitemOpen
  \bibfield  {author} {\bibinfo {author} {\bibfnamefont {S.}~\bibnamefont
  {Fern{\'{a}}ndez-Lorenzo}}, \bibinfo {author} {\bibfnamefont
  {D.}~\bibnamefont {Porras}},\ and\ \bibinfo {author} {\bibfnamefont {J.~J.}\
  \bibnamefont {Garc{\'{\i}}a-Ripoll}},\ }\bibfield  {title} {\bibinfo {title}
  {Hybrid quantum{\textendash}classical optimization with cardinality
  constraints and applications to finance},\ }\href
  {https://doi.org/10.1088/2058-9565/abf9af} {\bibfield  {journal} {\bibinfo
  {journal} {Quantum Sci. Technol.}\ }\textbf {\bibinfo {volume} {6}},\
  \bibinfo {pages} {034010} (\bibinfo {year} {2021})}\BibitemShut {NoStop}%
\bibitem [{\citenamefont {Saleem}\ \emph {et~al.}(2021)\citenamefont {Saleem},
  \citenamefont {Tomesh}, \citenamefont {Perlin}, \citenamefont {Gokhale},\
  and\ \citenamefont {Suchara}}]{Saleem2021}%
  \BibitemOpen
  \bibfield  {author} {\bibinfo {author} {\bibfnamefont {Z.~H.}\ \bibnamefont
  {Saleem}}, \bibinfo {author} {\bibfnamefont {T.}~\bibnamefont {Tomesh}},
  \bibinfo {author} {\bibfnamefont {M.~A.}\ \bibnamefont {Perlin}}, \bibinfo
  {author} {\bibfnamefont {P.}~\bibnamefont {Gokhale}},\ and\ \bibinfo {author}
  {\bibfnamefont {M.}~\bibnamefont {Suchara}},\ }\href@noop {} {\bibinfo
  {title} {Quantum divide and conquer for combinatorial optimization and
  distributed computing}} (\bibinfo {year} {2021}),\ \Eprint
  {https://arxiv.org/abs/2107.07532} {arXiv:2107.07532 [quant-ph]} \BibitemShut
  {NoStop}%
\bibitem [{\citenamefont {Kochenberger}\ \emph {et~al.}(2014)\citenamefont
  {Kochenberger}, \citenamefont {Hao}, \citenamefont {Glover}, \citenamefont
  {Lewis}, \citenamefont {L{\"u}}, \citenamefont {Wang},\ and\ \citenamefont
  {Wang}}]{Kochenberger2014}%
  \BibitemOpen
  \bibfield  {author} {\bibinfo {author} {\bibfnamefont {G.}~\bibnamefont
  {Kochenberger}}, \bibinfo {author} {\bibfnamefont {J.-K.}\ \bibnamefont
  {Hao}}, \bibinfo {author} {\bibfnamefont {F.}~\bibnamefont {Glover}},
  \bibinfo {author} {\bibfnamefont {M.}~\bibnamefont {Lewis}}, \bibinfo
  {author} {\bibfnamefont {Z.}~\bibnamefont {L{\"u}}}, \bibinfo {author}
  {\bibfnamefont {H.}~\bibnamefont {Wang}},\ and\ \bibinfo {author}
  {\bibfnamefont {Y.}~\bibnamefont {Wang}},\ }\bibfield  {title} {\bibinfo
  {title} {The unconstrained binary quadratic programming problem: a survey},\
  }\href {https://doi.org/10.1007/s10878-014-9734-0} {\bibfield  {journal}
  {\bibinfo  {journal} {J. Comb. Optim.}\ }\textbf {\bibinfo {volume} {28}},\
  \bibinfo {pages} {58} (\bibinfo {year} {2014})}\BibitemShut {NoStop}%
\bibitem [{\citenamefont {Lucas}(2014)}]{Lu14}%
  \BibitemOpen
  \bibfield  {author} {\bibinfo {author} {\bibfnamefont {A.}~\bibnamefont
  {Lucas}},\ }\bibfield  {title} {\bibinfo {title} {Ising formulations of many
  {NP} problems},\ }\href {https://doi.org/10.3389/fphy.2014.00005} {\bibfield
  {journal} {\bibinfo  {journal} {Front. Phys.}\ }\textbf {\bibinfo {volume}
  {2}},\ \bibinfo {pages} {5} (\bibinfo {year} {2014})}\BibitemShut {NoStop}%
\bibitem [{\citenamefont {Glover}\ \emph {et~al.}(2019)\citenamefont {Glover},
  \citenamefont {Kochenberger},\ and\ \citenamefont {Du}}]{Glover2019}%
  \BibitemOpen
  \bibfield  {author} {\bibinfo {author} {\bibfnamefont {F.}~\bibnamefont
  {Glover}}, \bibinfo {author} {\bibfnamefont {G.}~\bibnamefont
  {Kochenberger}},\ and\ \bibinfo {author} {\bibfnamefont {Y.}~\bibnamefont
  {Du}},\ }\bibfield  {title} {\bibinfo {title} {Quantum {B}ridge {A}nalytics
  {I}: a tutorial on formulating and using {QUBO} models},\ }\href
  {https://link.springer.com/article/10.1007/s10288-019-00424-y#citeas}
  {\bibfield  {journal} {\bibinfo  {journal} {4OR}\ }\textbf {\bibinfo {volume}
  {17}},\ \bibinfo {pages} {335} (\bibinfo {year} {2019})}\BibitemShut
  {NoStop}%
\bibitem [{\citenamefont {Peruzzo}\ \emph {et~al.}(2014)\citenamefont
  {Peruzzo}, \citenamefont {McClean}, \citenamefont {Shadbolt}, \citenamefont
  {Yung}, \citenamefont {Zhou}, \citenamefont {Love}, \citenamefont
  {Aspuru-Guzik},\ and\ \citenamefont {O'Brien}}]{Peruzzo2014}%
  \BibitemOpen
  \bibfield  {author} {\bibinfo {author} {\bibfnamefont {A.}~\bibnamefont
  {Peruzzo}}, \bibinfo {author} {\bibfnamefont {J.}~\bibnamefont {McClean}},
  \bibinfo {author} {\bibfnamefont {P.}~\bibnamefont {Shadbolt}}, \bibinfo
  {author} {\bibfnamefont {M.-H.}\ \bibnamefont {Yung}}, \bibinfo {author}
  {\bibfnamefont {X.-Q.}\ \bibnamefont {Zhou}}, \bibinfo {author}
  {\bibfnamefont {P.~J.}\ \bibnamefont {Love}}, \bibinfo {author}
  {\bibfnamefont {A.}~\bibnamefont {Aspuru-Guzik}},\ and\ \bibinfo {author}
  {\bibfnamefont {J.~L.}\ \bibnamefont {O'Brien}},\ }\bibfield  {title}
  {\bibinfo {title} {A variational eigenvalue solver on a photonic quantum
  processor},\ }\href {https://doi.org/10.1038/ncomms5213} {\bibfield
  {journal} {\bibinfo  {journal} {Nat. Commun.}\ }\textbf {\bibinfo {volume}
  {5}},\ \bibinfo {pages} {4213} (\bibinfo {year} {2014})}\BibitemShut
  {NoStop}%
\bibitem [{\citenamefont {Farhi}\ \emph {et~al.}(2014)\citenamefont {Farhi},
  \citenamefont {Goldstone},\ and\ \citenamefont {Gutmann}}]{Farhi2014}%
  \BibitemOpen
  \bibfield  {author} {\bibinfo {author} {\bibfnamefont {E.}~\bibnamefont
  {Farhi}}, \bibinfo {author} {\bibfnamefont {J.}~\bibnamefont {Goldstone}},\
  and\ \bibinfo {author} {\bibfnamefont {S.}~\bibnamefont {Gutmann}},\
  }\href@noop {} {\bibinfo {title} {A quantum approximate optimization
  algorithm}} (\bibinfo {year} {2014}),\ \Eprint
  {https://arxiv.org/abs/1411.4028} {arXiv:1411.4028 [quant-ph]} \BibitemShut
  {NoStop}%
\bibitem [{\citenamefont {Moll}\ \emph {et~al.}(2018)\citenamefont {Moll},
  \citenamefont {Barkoutsos}, \citenamefont {Bishop}, \citenamefont {Chow},
  \citenamefont {Cross}, \citenamefont {Egger}, \citenamefont {Filipp},
  \citenamefont {Fuhrer}, \citenamefont {Gambetta}, \citenamefont {Ganzhorn},\
  and\ \citenamefont {et~al.}}]{Moll2018}%
  \BibitemOpen
  \bibfield  {author} {\bibinfo {author} {\bibfnamefont {N.}~\bibnamefont
  {Moll}}, \bibinfo {author} {\bibfnamefont {P.}~\bibnamefont {Barkoutsos}},
  \bibinfo {author} {\bibfnamefont {L.~S.}\ \bibnamefont {Bishop}}, \bibinfo
  {author} {\bibfnamefont {J.~M.}\ \bibnamefont {Chow}}, \bibinfo {author}
  {\bibfnamefont {A.}~\bibnamefont {Cross}}, \bibinfo {author} {\bibfnamefont
  {D.~J.}\ \bibnamefont {Egger}}, \bibinfo {author} {\bibfnamefont
  {S.}~\bibnamefont {Filipp}}, \bibinfo {author} {\bibfnamefont
  {A.}~\bibnamefont {Fuhrer}}, \bibinfo {author} {\bibfnamefont {J.~M.}\
  \bibnamefont {Gambetta}}, \bibinfo {author} {\bibfnamefont {M.}~\bibnamefont
  {Ganzhorn}},\ and\ \bibinfo {author} {\bibnamefont {et~al.}},\ }\bibfield
  {title} {\bibinfo {title} {Quantum optimization using variational algorithms
  on near-term quantum devices},\ }\href
  {https://doi.org/10.1088/2058-9565/aab822} {\bibfield  {journal} {\bibinfo
  {journal} {Quantum Sci. Technol.}\ }\textbf {\bibinfo {volume} {3}},\
  \bibinfo {pages} {030503} (\bibinfo {year} {2018})}\BibitemShut {NoStop}%
\bibitem [{\citenamefont {Bravo-Prieto}\ \emph
  {et~al.}(2020{\natexlab{b}})\citenamefont {Bravo-Prieto}, \citenamefont
  {Lumbreras-Zarapico}, \citenamefont {Tagliacozzo},\ and\ \citenamefont
  {Latorre}}]{Prieto2020VQE}%
  \BibitemOpen
  \bibfield  {author} {\bibinfo {author} {\bibfnamefont {C.}~\bibnamefont
  {Bravo-Prieto}}, \bibinfo {author} {\bibfnamefont {J.}~\bibnamefont
  {Lumbreras-Zarapico}}, \bibinfo {author} {\bibfnamefont {L.}~\bibnamefont
  {Tagliacozzo}},\ and\ \bibinfo {author} {\bibfnamefont {J.~I.}\ \bibnamefont
  {Latorre}},\ }\bibfield  {title} {\bibinfo {title} {Scaling of variational
  quantum circuit depth for condensed matter systems},\ }\href
  {https://doi.org/10.22331/q-2020-05-28-272} {\bibfield  {journal} {\bibinfo
  {journal} {{Quantum}}\ }\textbf {\bibinfo {volume} {4}},\ \bibinfo {pages}
  {272} (\bibinfo {year} {2020}{\natexlab{b}})}\BibitemShut {NoStop}%
\bibitem [{\citenamefont {Garcia-Saez}\ and\ \citenamefont
  {Latorre}(2018)}]{Saez2018}%
  \BibitemOpen
  \bibfield  {author} {\bibinfo {author} {\bibfnamefont {A.}~\bibnamefont
  {Garcia-Saez}}\ and\ \bibinfo {author} {\bibfnamefont {J.~I.}\ \bibnamefont
  {Latorre}},\ }\href@noop {} {\bibinfo {title} {Addressing hard classical
  problems with adiabatically assisted variational quantum eigensolvers}}
  (\bibinfo {year} {2018}),\ \Eprint {https://arxiv.org/abs/1806.02287}
  {arXiv:1806.02287 [quant-ph]} \BibitemShut {NoStop}%
\bibitem [{\citenamefont {D\'{\i}ez-Valle}\ \emph {et~al.}(2021)\citenamefont
  {D\'{\i}ez-Valle}, \citenamefont {Porras},\ and\ \citenamefont
  {Garc\'{\i}a-Ripoll}}]{Diez2021}%
  \BibitemOpen
  \bibfield  {author} {\bibinfo {author} {\bibfnamefont {P.}~\bibnamefont
  {D\'{\i}ez-Valle}}, \bibinfo {author} {\bibfnamefont {D.}~\bibnamefont
  {Porras}},\ and\ \bibinfo {author} {\bibfnamefont {J.~J.}\ \bibnamefont
  {Garc\'{\i}a-Ripoll}},\ }\bibfield  {title} {\bibinfo {title} {{Quantum
  variational optimization: The role of entanglement and problem hardness}},\
  }\href {https://doi.org/10.1103/PhysRevA.104.062426} {\bibfield  {journal}
  {\bibinfo  {journal} {Phys. Rev. A}\ }\textbf {\bibinfo {volume} {104}},\
  \bibinfo {pages} {062426} (\bibinfo {year} {2021})}\BibitemShut {NoStop}%
\bibitem [{\citenamefont {Farhi}\ \emph {et~al.}(2001)\citenamefont {Farhi},
  \citenamefont {Goldstone}, \citenamefont {Gutmann}, \citenamefont {Lapan},
  \citenamefont {Lundgren},\ and\ \citenamefont {Preda}}]{FaEtAl01}%
  \BibitemOpen
  \bibfield  {author} {\bibinfo {author} {\bibfnamefont {E.}~\bibnamefont
  {Farhi}}, \bibinfo {author} {\bibfnamefont {J.}~\bibnamefont {Goldstone}},
  \bibinfo {author} {\bibfnamefont {S.}~\bibnamefont {Gutmann}}, \bibinfo
  {author} {\bibfnamefont {J.}~\bibnamefont {Lapan}}, \bibinfo {author}
  {\bibfnamefont {A.}~\bibnamefont {Lundgren}},\ and\ \bibinfo {author}
  {\bibfnamefont {D.}~\bibnamefont {Preda}},\ }\bibfield  {title} {\bibinfo
  {title} {A quantum adiabatic evolution algorithm applied to random instances
  of an {NP}-complete problem},\ }\href
  {https://doi.org/10.1126/science.1057726} {\bibfield  {journal} {\bibinfo
  {journal} {Science}\ }\textbf {\bibinfo {volume} {292}},\ \bibinfo {pages}
  {472} (\bibinfo {year} {2001})}\BibitemShut {NoStop}%
\bibitem [{\citenamefont {Kadowaki}\ and\ \citenamefont
  {Nishimori}(1998)}]{TaHi98}%
  \BibitemOpen
  \bibfield  {author} {\bibinfo {author} {\bibfnamefont {T.}~\bibnamefont
  {Kadowaki}}\ and\ \bibinfo {author} {\bibfnamefont {H.}~\bibnamefont
  {Nishimori}},\ }\bibfield  {title} {\bibinfo {title} {Quantum annealing in
  the transverse {I}sing model},\ }\href
  {https://doi.org/10.1103/PhysRevE.58.5355} {\bibfield  {journal} {\bibinfo
  {journal} {Phys. Rev. E}\ }\textbf {\bibinfo {volume} {58}},\ \bibinfo
  {pages} {5355} (\bibinfo {year} {1998})}\BibitemShut {NoStop}%
\bibitem [{\citenamefont {Farhi}\ and\ \citenamefont {Harrow}(2019)}]{FaHa19}%
  \BibitemOpen
  \bibfield  {author} {\bibinfo {author} {\bibfnamefont {E.}~\bibnamefont
  {Farhi}}\ and\ \bibinfo {author} {\bibfnamefont {A.~W.}\ \bibnamefont
  {Harrow}},\ }\href@noop {} {\bibinfo {title} {Quantum supremacy through the
  quantum approximate optimization algorithm}} (\bibinfo {year} {2019}),\
  \Eprint {https://arxiv.org/abs/1602.07674} {arXiv:1602.07674 [quant-ph]}
  \BibitemShut {NoStop}%
\bibitem [{\citenamefont {Zhou}\ \emph {et~al.}(2020)\citenamefont {Zhou},
  \citenamefont {Wang}, \citenamefont {Choi}, \citenamefont {Pichler},\ and\
  \citenamefont {Lukin}}]{Zhou2020}%
  \BibitemOpen
  \bibfield  {author} {\bibinfo {author} {\bibfnamefont {L.}~\bibnamefont
  {Zhou}}, \bibinfo {author} {\bibfnamefont {S.-T.}\ \bibnamefont {Wang}},
  \bibinfo {author} {\bibfnamefont {S.}~\bibnamefont {Choi}}, \bibinfo {author}
  {\bibfnamefont {H.}~\bibnamefont {Pichler}},\ and\ \bibinfo {author}
  {\bibfnamefont {M.~D.}\ \bibnamefont {Lukin}},\ }\bibfield  {title} {\bibinfo
  {title} {Quantum approximate optimization algorithm: Performance, mechanism,
  and implementation on near-term devices},\ }\href
  {https://doi.org/10.1103/PhysRevX.10.021067} {\bibfield  {journal} {\bibinfo
  {journal} {Phys. Rev. X}\ }\textbf {\bibinfo {volume} {10}},\ \bibinfo
  {pages} {021067} (\bibinfo {year} {2020})}\BibitemShut {NoStop}%
\bibitem [{\citenamefont {Moussa}\ \emph {et~al.}(2020)\citenamefont {Moussa},
  \citenamefont {Calandra},\ and\ \citenamefont {Dunjko}}]{Moussa2020}%
  \BibitemOpen
  \bibfield  {author} {\bibinfo {author} {\bibfnamefont {C.}~\bibnamefont
  {Moussa}}, \bibinfo {author} {\bibfnamefont {H.}~\bibnamefont {Calandra}},\
  and\ \bibinfo {author} {\bibfnamefont {V.}~\bibnamefont {Dunjko}},\
  }\bibfield  {title} {\bibinfo {title} {To quantum or not to quantum: towards
  algorithm selection in near-term quantum optimization},\ }\href
  {https://doi.org/10.1088/2058-9565/abb8e5} {\bibfield  {journal} {\bibinfo
  {journal} {Quantum Sci. Technol.}\ }\textbf {\bibinfo {volume} {5}},\
  \bibinfo {pages} {044009} (\bibinfo {year} {2020})}\BibitemShut {NoStop}%
\bibitem [{\citenamefont {Harrigan}\ \emph {et~al.}(2021)\citenamefont
  {Harrigan}, \citenamefont {Sung}, \citenamefont {Neeley}, \citenamefont
  {Satzinger}, \citenamefont {Arute}, \citenamefont {Arya}, \citenamefont
  {Atalaya}, \citenamefont {Bardin}, \citenamefont {Barends}, \citenamefont
  {Boixo},\ and\ \citenamefont {et~al.}}]{HaEtAl21}%
  \BibitemOpen
  \bibfield  {author} {\bibinfo {author} {\bibfnamefont {M.~P.}\ \bibnamefont
  {Harrigan}}, \bibinfo {author} {\bibfnamefont {K.~J.}\ \bibnamefont {Sung}},
  \bibinfo {author} {\bibfnamefont {M.}~\bibnamefont {Neeley}}, \bibinfo
  {author} {\bibfnamefont {K.~J.}\ \bibnamefont {Satzinger}}, \bibinfo {author}
  {\bibfnamefont {F.}~\bibnamefont {Arute}}, \bibinfo {author} {\bibfnamefont
  {K.}~\bibnamefont {Arya}}, \bibinfo {author} {\bibfnamefont {J.}~\bibnamefont
  {Atalaya}}, \bibinfo {author} {\bibfnamefont {J.~C.}\ \bibnamefont {Bardin}},
  \bibinfo {author} {\bibfnamefont {R.}~\bibnamefont {Barends}}, \bibinfo
  {author} {\bibfnamefont {S.}~\bibnamefont {Boixo}},\ and\ \bibinfo {author}
  {\bibnamefont {et~al.}},\ }\bibfield  {title} {\bibinfo {title} {Quantum
  approximate optimization of non-planar graph problems on a planar
  superconducting processor},\ }\href
  {https://doi.org/10.1038/s41567-020-01105-y} {\bibfield  {journal} {\bibinfo
  {journal} {Nat. Phys.}\ }\textbf {\bibinfo {volume} {17}},\ \bibinfo {pages}
  {332} (\bibinfo {year} {2021})}\BibitemShut {NoStop}%
\bibitem [{\citenamefont {Benedetti}\ \emph {et~al.}(2021)\citenamefont
  {Benedetti}, \citenamefont {Fiorentini},\ and\ \citenamefont
  {Lubasch}}]{BeFiLu20}%
  \BibitemOpen
  \bibfield  {author} {\bibinfo {author} {\bibfnamefont {M.}~\bibnamefont
  {Benedetti}}, \bibinfo {author} {\bibfnamefont {M.}~\bibnamefont
  {Fiorentini}},\ and\ \bibinfo {author} {\bibfnamefont {M.}~\bibnamefont
  {Lubasch}},\ }\bibfield  {title} {\bibinfo {title} {Hardware-efficient
  variational quantum algorithms for time evolution},\ }\href
  {https://doi.org/10.1103/PhysRevResearch.3.033083} {\bibfield  {journal}
  {\bibinfo  {journal} {Phys. Rev. Research}\ }\textbf {\bibinfo {volume}
  {3}},\ \bibinfo {pages} {033083} (\bibinfo {year} {2021})}\BibitemShut
  {NoStop}%
\bibitem [{\citenamefont {H\r{a}stad}(2001)}]{Hastad2001}%
  \BibitemOpen
  \bibfield  {author} {\bibinfo {author} {\bibfnamefont {J.}~\bibnamefont
  {H\r{a}stad}},\ }\bibfield  {title} {\bibinfo {title} {Some optimal
  inapproximability results},\ }\href {https://doi.org/10.1145/502090.502098}
  {\bibfield  {journal} {\bibinfo  {journal} {J. ACM}\ }\textbf {\bibinfo
  {volume} {48}},\ \bibinfo {pages} {798–859} (\bibinfo {year}
  {2001})}\BibitemShut {NoStop}%
\bibitem [{\citenamefont {Berman}\ and\ \citenamefont
  {Karpinski}(1999)}]{Berman2002}%
  \BibitemOpen
  \bibfield  {author} {\bibinfo {author} {\bibfnamefont {P.}~\bibnamefont
  {Berman}}\ and\ \bibinfo {author} {\bibfnamefont {M.}~\bibnamefont
  {Karpinski}},\ }\bibfield  {title} {\bibinfo {title} {On some tighter
  inapproximability results (extended abstract)},\ }in\ \href
  {https://doi.org/https://doi.org/10.1007/3-540-48523-6_17} {\emph {\bibinfo
  {booktitle} {Automata, Languages and Programming}}},\ \bibinfo {editor}
  {edited by\ \bibinfo {editor} {\bibfnamefont {J.}~\bibnamefont {Wiedermann}},
  \bibinfo {editor} {\bibfnamefont {P.}~\bibnamefont {van Emde~Boas}},\ and\
  \bibinfo {editor} {\bibfnamefont {M.}~\bibnamefont {Nielsen}}}\ (\bibinfo
  {publisher} {Springer Berlin Heidelberg},\ \bibinfo {address} {Berlin,
  Heidelberg},\ \bibinfo {year} {1999})\ pp.\ \bibinfo {pages}
  {200--209}\BibitemShut {NoStop}%
\bibitem [{\citenamefont {Amaro}\ \emph {et~al.}(2021)\citenamefont {Amaro},
  \citenamefont {Rosenkranz}, \citenamefont {Fitzpatrick}, \citenamefont
  {Hirano},\ and\ \citenamefont {Fiorentini}}]{AmEtAl21}%
  \BibitemOpen
  \bibfield  {author} {\bibinfo {author} {\bibfnamefont {D.}~\bibnamefont
  {Amaro}}, \bibinfo {author} {\bibfnamefont {M.}~\bibnamefont {Rosenkranz}},
  \bibinfo {author} {\bibfnamefont {N.}~\bibnamefont {Fitzpatrick}}, \bibinfo
  {author} {\bibfnamefont {K.}~\bibnamefont {Hirano}},\ and\ \bibinfo {author}
  {\bibfnamefont {M.}~\bibnamefont {Fiorentini}},\ }\href@noop {} {\bibinfo
  {title} {A case study of variational quantum algorithms for a job shop
  scheduling problem}} (\bibinfo {year} {2021}),\ \Eprint
  {https://arxiv.org/abs/2109.03745} {arXiv:2109.03745 [quant-ph]} \BibitemShut
  {NoStop}%
\bibitem [{\citenamefont {Wei\ss{}e}\ \emph {et~al.}(2006)\citenamefont
  {Wei\ss{}e}, \citenamefont {Wellein}, \citenamefont {Alvermann},\ and\
  \citenamefont {Fehske}}]{Weiss2006}%
  \BibitemOpen
  \bibfield  {author} {\bibinfo {author} {\bibfnamefont {A.}~\bibnamefont
  {Wei\ss{}e}}, \bibinfo {author} {\bibfnamefont {G.}~\bibnamefont {Wellein}},
  \bibinfo {author} {\bibfnamefont {A.}~\bibnamefont {Alvermann}},\ and\
  \bibinfo {author} {\bibfnamefont {H.}~\bibnamefont {Fehske}},\ }\bibfield
  {title} {\bibinfo {title} {The kernel polynomial method},\ }\href
  {https://doi.org/10.1103/RevModPhys.78.275} {\bibfield  {journal} {\bibinfo
  {journal} {Rev. Mod. Phys.}\ }\textbf {\bibinfo {volume} {78}},\ \bibinfo
  {pages} {275} (\bibinfo {year} {2006})}\BibitemShut {NoStop}%
\bibitem [{\citenamefont {Trefethen}\ and\ \citenamefont {Bau}(1997)}]{TrBa97}%
  \BibitemOpen
  \bibfield  {author} {\bibinfo {author} {\bibfnamefont {L.~N.}\ \bibnamefont
  {Trefethen}}\ and\ \bibinfo {author} {\bibfnamefont {D.}~\bibnamefont
  {Bau}},\ }\href
  {https://my.siam.org/Store/Product/viewproduct/?ProductId=950} {\emph
  {\bibinfo {title} {{Numerical linear algebra}}}}\ (\bibinfo  {publisher}
  {SIAM (Society for Industrial and Applied Mathematics) Philadelphia},\
  \bibinfo {year} {1997})\BibitemShut {NoStop}%
\bibitem [{\citenamefont {Noble}\ \emph {et~al.}(2013)\citenamefont {Noble},
  \citenamefont {Lubasch},\ and\ \citenamefont {Jentschura}}]{NoLuJe13}%
  \BibitemOpen
  \bibfield  {author} {\bibinfo {author} {\bibfnamefont {J.}~\bibnamefont
  {Noble}}, \bibinfo {author} {\bibfnamefont {M.}~\bibnamefont {Lubasch}},\
  and\ \bibinfo {author} {\bibfnamefont {U.}~\bibnamefont {Jentschura}},\
  }\bibfield  {title} {\bibinfo {title} {{Generalized Householder}
  transformations for the complex symmetric eigenvalue problems},\ }\href
  {https://doi.org/https://doi.org/10.1140/epjp/i2013-13093-1} {\bibfield
  {journal} {\bibinfo  {journal} {Eur. Phys. J. Plus}\ }\textbf {\bibinfo
  {volume} {128}},\ \bibinfo {pages} {93} (\bibinfo {year} {2013})}\BibitemShut
  {NoStop}%
\bibitem [{\citenamefont {Noble}\ \emph {et~al.}(2017)\citenamefont {Noble},
  \citenamefont {Lubasch}, \citenamefont {Stevens},\ and\ \citenamefont
  {Jentschura}}]{NoEtAl17}%
  \BibitemOpen
  \bibfield  {author} {\bibinfo {author} {\bibfnamefont {J.}~\bibnamefont
  {Noble}}, \bibinfo {author} {\bibfnamefont {M.}~\bibnamefont {Lubasch}},
  \bibinfo {author} {\bibfnamefont {J.}~\bibnamefont {Stevens}},\ and\ \bibinfo
  {author} {\bibfnamefont {U.}~\bibnamefont {Jentschura}},\ }\bibfield  {title}
  {\bibinfo {title} {Diagonalization of complex symmetric matrices:
  {Generalized Householder} reflections, iterative deflation and implicit
  shifts},\ }\href {https://doi.org/https://doi.org/10.1016/j.cpc.2017.06.014}
  {\bibfield  {journal} {\bibinfo  {journal} {Comput. Phys. Commun.}\ }\textbf
  {\bibinfo {volume} {221}},\ \bibinfo {pages} {304} (\bibinfo {year}
  {2017})}\BibitemShut {NoStop}%
\bibitem [{\citenamefont {Ge}\ \emph {et~al.}(2019)\citenamefont {Ge},
  \citenamefont {Tura},\ and\ \citenamefont {Cirac}}]{ge2018faster}%
  \BibitemOpen
  \bibfield  {author} {\bibinfo {author} {\bibfnamefont {Y.}~\bibnamefont
  {Ge}}, \bibinfo {author} {\bibfnamefont {J.}~\bibnamefont {Tura}},\ and\
  \bibinfo {author} {\bibfnamefont {J.~I.}\ \bibnamefont {Cirac}},\ }\bibfield
  {title} {\bibinfo {title} {Faster ground state preparation and high-precision
  ground energy estimation with fewer qubits},\ }\href
  {https://doi.org/10.1063/1.5027484} {\bibfield  {journal} {\bibinfo
  {journal} {J. Math. Phys.}\ }\textbf {\bibinfo {volume} {60}},\ \bibinfo
  {pages} {022202} (\bibinfo {year} {2019})}\BibitemShut {NoStop}%
\bibitem [{\citenamefont {Lu}\ \emph {et~al.}(2020)\citenamefont {Lu},
  \citenamefont {Bañuls},\ and\ \citenamefont {Cirac}}]{LuBaCi20}%
  \BibitemOpen
  \bibfield  {author} {\bibinfo {author} {\bibfnamefont {S.}~\bibnamefont
  {Lu}}, \bibinfo {author} {\bibfnamefont {M.~C.}\ \bibnamefont {Bañuls}},\
  and\ \bibinfo {author} {\bibfnamefont {J.~I.}\ \bibnamefont {Cirac}},\
  }\href@noop {} {\bibinfo {title} {Algorithms for quantum simulation at finite
  energies}} (\bibinfo {year} {2020}),\ \Eprint
  {https://arxiv.org/abs/2006.03032} {arXiv:2006.03032 [quant-ph]} \BibitemShut
  {NoStop}%
\bibitem [{\citenamefont {Yang}\ \emph {et~al.}(2020)\citenamefont {Yang},
  \citenamefont {Iblisdir}, \citenamefont {Cirac},\ and\ \citenamefont
  {Ba\~nuls}}]{YaEtAl20}%
  \BibitemOpen
  \bibfield  {author} {\bibinfo {author} {\bibfnamefont {Y.}~\bibnamefont
  {Yang}}, \bibinfo {author} {\bibfnamefont {S.}~\bibnamefont {Iblisdir}},
  \bibinfo {author} {\bibfnamefont {J.~I.}\ \bibnamefont {Cirac}},\ and\
  \bibinfo {author} {\bibfnamefont {M.~C.}\ \bibnamefont {Ba\~nuls}},\
  }\bibfield  {title} {\bibinfo {title} {Probing thermalization through
  spectral analysis with matrix product operators},\ }\href
  {https://doi.org/10.1103/PhysRevLett.124.100602} {\bibfield  {journal}
  {\bibinfo  {journal} {Phys. Rev. Lett.}\ }\textbf {\bibinfo {volume} {124}},\
  \bibinfo {pages} {100602} (\bibinfo {year} {2020})}\BibitemShut {NoStop}%
\bibitem [{\citenamefont {Ba\~nuls}\ \emph {et~al.}(2020)\citenamefont
  {Ba\~nuls}, \citenamefont {Huse},\ and\ \citenamefont {Cirac}}]{BaHuCi20}%
  \BibitemOpen
  \bibfield  {author} {\bibinfo {author} {\bibfnamefont {M.~C.}\ \bibnamefont
  {Ba\~nuls}}, \bibinfo {author} {\bibfnamefont {D.~A.}\ \bibnamefont {Huse}},\
  and\ \bibinfo {author} {\bibfnamefont {J.~I.}\ \bibnamefont {Cirac}},\
  }\bibfield  {title} {\bibinfo {title} {Entanglement and its relation to
  energy variance for local one-dimensional {Hamiltonians}},\ }\href
  {https://doi.org/10.1103/PhysRevB.101.144305} {\bibfield  {journal} {\bibinfo
   {journal} {Phys. Rev. B}\ }\textbf {\bibinfo {volume} {101}},\ \bibinfo
  {pages} {144305} (\bibinfo {year} {2020})}\BibitemShut {NoStop}%
\bibitem [{\citenamefont {\ifmmode~\mbox{\c{C}}\else \c{C}\fi{}akan}\ \emph
  {et~al.}(2021)\citenamefont {\ifmmode~\mbox{\c{C}}\else \c{C}\fi{}akan},
  \citenamefont {Cirac},\ and\ \citenamefont {Ba\~nuls}}]{CaCiBa21}%
  \BibitemOpen
  \bibfield  {author} {\bibinfo {author} {\bibfnamefont {A.}~\bibnamefont
  {\ifmmode~\mbox{\c{C}}\else \c{C}\fi{}akan}}, \bibinfo {author}
  {\bibfnamefont {J.~I.}\ \bibnamefont {Cirac}},\ and\ \bibinfo {author}
  {\bibfnamefont {M.~C.}\ \bibnamefont {Ba\~nuls}},\ }\bibfield  {title}
  {\bibinfo {title} {Approximating the long time average of the density
  operator: Diagonal ensemble},\ }\href
  {https://doi.org/10.1103/PhysRevB.103.115113} {\bibfield  {journal} {\bibinfo
   {journal} {Phys. Rev. B}\ }\textbf {\bibinfo {volume} {103}},\ \bibinfo
  {pages} {115113} (\bibinfo {year} {2021})}\BibitemShut {NoStop}%
\bibitem [{\citenamefont {Mitarai}\ \emph {et~al.}(2018)\citenamefont
  {Mitarai}, \citenamefont {Negoro}, \citenamefont {Kitagawa},\ and\
  \citenamefont {Fujii}}]{Mitarai2018}%
  \BibitemOpen
  \bibfield  {author} {\bibinfo {author} {\bibfnamefont {K.}~\bibnamefont
  {Mitarai}}, \bibinfo {author} {\bibfnamefont {M.}~\bibnamefont {Negoro}},
  \bibinfo {author} {\bibfnamefont {M.}~\bibnamefont {Kitagawa}},\ and\
  \bibinfo {author} {\bibfnamefont {K.}~\bibnamefont {Fujii}},\ }\bibfield
  {title} {\bibinfo {title} {Quantum circuit learning},\ }\href
  {https://doi.org/10.1103/PhysRevA.98.032309} {\bibfield  {journal} {\bibinfo
  {journal} {Phys. Rev. A}\ }\textbf {\bibinfo {volume} {98}},\ \bibinfo
  {pages} {032309} (\bibinfo {year} {2018})}\BibitemShut {NoStop}%
\bibitem [{\citenamefont {Schuld}\ \emph {et~al.}(2019)\citenamefont {Schuld},
  \citenamefont {Bergholm}, \citenamefont {Gogolin}, \citenamefont {Izaac},\
  and\ \citenamefont {Killoran}}]{Schuld2019}%
  \BibitemOpen
  \bibfield  {author} {\bibinfo {author} {\bibfnamefont {M.}~\bibnamefont
  {Schuld}}, \bibinfo {author} {\bibfnamefont {V.}~\bibnamefont {Bergholm}},
  \bibinfo {author} {\bibfnamefont {C.}~\bibnamefont {Gogolin}}, \bibinfo
  {author} {\bibfnamefont {J.}~\bibnamefont {Izaac}},\ and\ \bibinfo {author}
  {\bibfnamefont {N.}~\bibnamefont {Killoran}},\ }\bibfield  {title} {\bibinfo
  {title} {Evaluating analytic gradients on quantum hardware},\ }\href
  {https://doi.org/10.1103/PhysRevA.99.032331} {\bibfield  {journal} {\bibinfo
  {journal} {Phys. Rev. A}\ }\textbf {\bibinfo {volume} {99}},\ \bibinfo
  {pages} {032331} (\bibinfo {year} {2019})}\BibitemShut {NoStop}%
\bibitem [{\citenamefont {Vidal}(2008)}]{Vi08}%
  \BibitemOpen
  \bibfield  {author} {\bibinfo {author} {\bibfnamefont {G.}~\bibnamefont
  {Vidal}},\ }\bibfield  {title} {\bibinfo {title} {Class of quantum many-body
  states that can be efficiently simulated},\ }\href
  {https://doi.org/10.1103/PhysRevLett.101.110501} {\bibfield  {journal}
  {\bibinfo  {journal} {Phys. Rev. Lett.}\ }\textbf {\bibinfo {volume} {101}},\
  \bibinfo {pages} {110501} (\bibinfo {year} {2008})}\BibitemShut {NoStop}%
\bibitem [{\citenamefont {Pino}\ \emph {et~al.}(2021)\citenamefont {Pino},
  \citenamefont {Dreiling}, \citenamefont {Figgatt}, \citenamefont {Gaebler},
  \citenamefont {Moses}, \citenamefont {Allman}, \citenamefont {Baldwin},
  \citenamefont {Foss-Feig}, \citenamefont {Hayes}, \citenamefont {Mayer},
  \citenamefont {Ryan-Anderson},\ and\ \citenamefont {Neyenhuis}}]{Pino2021}%
  \BibitemOpen
  \bibfield  {author} {\bibinfo {author} {\bibfnamefont {J.~M.}\ \bibnamefont
  {Pino}}, \bibinfo {author} {\bibfnamefont {J.~M.}\ \bibnamefont {Dreiling}},
  \bibinfo {author} {\bibfnamefont {C.}~\bibnamefont {Figgatt}}, \bibinfo
  {author} {\bibfnamefont {J.~P.}\ \bibnamefont {Gaebler}}, \bibinfo {author}
  {\bibfnamefont {S.~A.}\ \bibnamefont {Moses}}, \bibinfo {author}
  {\bibfnamefont {M.~S.}\ \bibnamefont {Allman}}, \bibinfo {author}
  {\bibfnamefont {C.~H.}\ \bibnamefont {Baldwin}}, \bibinfo {author}
  {\bibfnamefont {M.}~\bibnamefont {Foss-Feig}}, \bibinfo {author}
  {\bibfnamefont {D.}~\bibnamefont {Hayes}}, \bibinfo {author} {\bibfnamefont
  {K.}~\bibnamefont {Mayer}}, \bibinfo {author} {\bibfnamefont
  {C.}~\bibnamefont {Ryan-Anderson}},\ and\ \bibinfo {author} {\bibfnamefont
  {B.}~\bibnamefont {Neyenhuis}},\ }\bibfield  {title} {\bibinfo {title}
  {Demonstration of the trapped-ion quantum {CCD} computer architecture},\
  }\href {https://doi.org/10.1038/s41586-021-03318-4} {\bibfield  {journal}
  {\bibinfo  {journal} {Nature}\ }\textbf {\bibinfo {volume} {592}},\ \bibinfo
  {pages} {209} (\bibinfo {year} {2021})}\BibitemShut {NoStop}%
\bibitem [{\citenamefont {Sivarajah}\ \emph {et~al.}(2020)\citenamefont
  {Sivarajah}, \citenamefont {Dilkes}, \citenamefont {Cowtan}, \citenamefont
  {Simmons}, \citenamefont {Edgington},\ and\ \citenamefont
  {Duncan}}]{Sivarajah2020}%
  \BibitemOpen
  \bibfield  {author} {\bibinfo {author} {\bibfnamefont {S.}~\bibnamefont
  {Sivarajah}}, \bibinfo {author} {\bibfnamefont {S.}~\bibnamefont {Dilkes}},
  \bibinfo {author} {\bibfnamefont {A.}~\bibnamefont {Cowtan}}, \bibinfo
  {author} {\bibfnamefont {W.}~\bibnamefont {Simmons}}, \bibinfo {author}
  {\bibfnamefont {A.}~\bibnamefont {Edgington}},\ and\ \bibinfo {author}
  {\bibfnamefont {R.}~\bibnamefont {Duncan}},\ }\bibfield  {title} {\bibinfo
  {title} {Tket: a retargetable compiler for {NISQ} devices},\ }\href
  {https://doi.org/10.1088/2058-9565/ab8e92} {\bibfield  {journal} {\bibinfo
  {journal} {Quantum Sci. Technol.}\ }\textbf {\bibinfo {volume} {6}},\
  \bibinfo {pages} {014003} (\bibinfo {year} {2020})}\BibitemShut {NoStop}%
\bibitem [{\citenamefont {Helmberg}(2000)}]{Helmberg2000}%
  \BibitemOpen
  \bibfield  {author} {\bibinfo {author} {\bibfnamefont {C.}~\bibnamefont
  {Helmberg}},\ }\href@noop {} {\bibinfo {title} {Semidefinite programming for
  combinatorial optimization}} (\bibinfo {year} {2000})\BibitemShut {NoStop}%
\bibitem [{\citenamefont {Dalgaard}\ \emph {et~al.}(2020)\citenamefont
  {Dalgaard}, \citenamefont {Motzoi}, \citenamefont {Jensen},\ and\
  \citenamefont {Sherson}}]{Dalgaard2020}%
  \BibitemOpen
  \bibfield  {author} {\bibinfo {author} {\bibfnamefont {M.}~\bibnamefont
  {Dalgaard}}, \bibinfo {author} {\bibfnamefont {F.}~\bibnamefont {Motzoi}},
  \bibinfo {author} {\bibfnamefont {J.~H.~M.}\ \bibnamefont {Jensen}},\ and\
  \bibinfo {author} {\bibfnamefont {J.}~\bibnamefont {Sherson}},\ }\bibfield
  {title} {\bibinfo {title} {Hessian-based optimization of constrained quantum
  control},\ }\href {https://doi.org/10.1103/PhysRevA.102.042612} {\bibfield
  {journal} {\bibinfo  {journal} {Phys. Rev. A}\ }\textbf {\bibinfo {volume}
  {102}},\ \bibinfo {pages} {042612} (\bibinfo {year} {2020})}\BibitemShut
  {NoStop}%
\bibitem [{\citenamefont {Mari}\ \emph {et~al.}(2021)\citenamefont {Mari},
  \citenamefont {Bromley},\ and\ \citenamefont {Killoran}}]{Andrea2021}%
  \BibitemOpen
  \bibfield  {author} {\bibinfo {author} {\bibfnamefont {A.}~\bibnamefont
  {Mari}}, \bibinfo {author} {\bibfnamefont {T.~R.}\ \bibnamefont {Bromley}},\
  and\ \bibinfo {author} {\bibfnamefont {N.}~\bibnamefont {Killoran}},\
  }\bibfield  {title} {\bibinfo {title} {Estimating the gradient and
  higher-order derivatives on quantum hardware},\ }\href
  {https://doi.org/10.1103/PhysRevA.103.012405} {\bibfield  {journal} {\bibinfo
   {journal} {Phys. Rev. A}\ }\textbf {\bibinfo {volume} {103}},\ \bibinfo
  {pages} {012405} (\bibinfo {year} {2021})}\BibitemShut {NoStop}%
\bibitem [{\citenamefont {Barkoutsos}\ \emph {et~al.}(2020)\citenamefont
  {Barkoutsos}, \citenamefont {Nannicini}, \citenamefont {Robert},
  \citenamefont {Tavernelli},\ and\ \citenamefont {Woerner}}]{Barkoutsos2020}%
  \BibitemOpen
  \bibfield  {author} {\bibinfo {author} {\bibfnamefont {P.~K.}\ \bibnamefont
  {Barkoutsos}}, \bibinfo {author} {\bibfnamefont {G.}~\bibnamefont
  {Nannicini}}, \bibinfo {author} {\bibfnamefont {A.}~\bibnamefont {Robert}},
  \bibinfo {author} {\bibfnamefont {I.}~\bibnamefont {Tavernelli}},\ and\
  \bibinfo {author} {\bibfnamefont {S.}~\bibnamefont {Woerner}},\ }\bibfield
  {title} {\bibinfo {title} {Improving {V}ariational {Q}uantum {O}ptimization
  using {CV}a{R}},\ }\href {https://doi.org/10.22331/q-2020-04-20-256}
  {\bibfield  {journal} {\bibinfo  {journal} {Quantum}\ }\textbf {\bibinfo
  {volume} {4}},\ \bibinfo {pages} {256} (\bibinfo {year} {2020})}\BibitemShut
  {NoStop}%
\bibitem [{\citenamefont {Kolotouros}\ and\ \citenamefont
  {Wallden}(2021)}]{Kolotouros2021}%
  \BibitemOpen
  \bibfield  {author} {\bibinfo {author} {\bibfnamefont {I.}~\bibnamefont
  {Kolotouros}}\ and\ \bibinfo {author} {\bibfnamefont {P.}~\bibnamefont
  {Wallden}},\ }\href@noop {} {\bibinfo {title} {An evolving objective function
  for improved variational quantum optimisation}} (\bibinfo {year} {2021}),\
  \Eprint {https://arxiv.org/abs/2105.11766} {arXiv:2105.11766 [quant-ph]}
  \BibitemShut {NoStop}%
\bibitem [{\citenamefont {Cerezo}\ \emph
  {et~al.}(2021{\natexlab{b}})\citenamefont {Cerezo}, \citenamefont {Sone},
  \citenamefont {Volkoff}, \citenamefont {Cincio},\ and\ \citenamefont
  {Coles}}]{Cerezo2021}%
  \BibitemOpen
  \bibfield  {author} {\bibinfo {author} {\bibfnamefont {M.}~\bibnamefont
  {Cerezo}}, \bibinfo {author} {\bibfnamefont {A.}~\bibnamefont {Sone}},
  \bibinfo {author} {\bibfnamefont {T.}~\bibnamefont {Volkoff}}, \bibinfo
  {author} {\bibfnamefont {L.}~\bibnamefont {Cincio}},\ and\ \bibinfo {author}
  {\bibfnamefont {P.~J.}\ \bibnamefont {Coles}},\ }\bibfield  {title} {\bibinfo
  {title} {Cost function dependent barren plateaus in shallow parametrized
  quantum circuits},\ }\href {https://doi.org/10.1038/s41467-021-21728-w}
  {\bibfield  {journal} {\bibinfo  {journal} {Nat. Commun.}\ }\textbf {\bibinfo
  {volume} {12}},\ \bibinfo {pages} {1791} (\bibinfo {year}
  {2021}{\natexlab{b}})}\BibitemShut {NoStop}%
\bibitem [{\citenamefont {Wecker}\ \emph {et~al.}(2015)\citenamefont {Wecker},
  \citenamefont {Hastings},\ and\ \citenamefont {Troyer}}]{WeHaTr15}%
  \BibitemOpen
  \bibfield  {author} {\bibinfo {author} {\bibfnamefont {D.}~\bibnamefont
  {Wecker}}, \bibinfo {author} {\bibfnamefont {M.~B.}\ \bibnamefont
  {Hastings}},\ and\ \bibinfo {author} {\bibfnamefont {M.}~\bibnamefont
  {Troyer}},\ }\bibfield  {title} {\bibinfo {title} {Progress towards practical
  quantum variational algorithms},\ }\href
  {https://doi.org/10.1103/PhysRevA.92.042303} {\bibfield  {journal} {\bibinfo
  {journal} {Phys. Rev. A}\ }\textbf {\bibinfo {volume} {92}},\ \bibinfo
  {pages} {042303} (\bibinfo {year} {2015})}\BibitemShut {NoStop}%
\bibitem [{\citenamefont {Wiersema}\ \emph {et~al.}(2020)\citenamefont
  {Wiersema}, \citenamefont {Zhou}, \citenamefont {de~Sereville}, \citenamefont
  {Carrasquilla}, \citenamefont {Kim},\ and\ \citenamefont
  {Yuen}}]{Wiersema2020}%
  \BibitemOpen
  \bibfield  {author} {\bibinfo {author} {\bibfnamefont {R.}~\bibnamefont
  {Wiersema}}, \bibinfo {author} {\bibfnamefont {C.}~\bibnamefont {Zhou}},
  \bibinfo {author} {\bibfnamefont {Y.}~\bibnamefont {de~Sereville}}, \bibinfo
  {author} {\bibfnamefont {J.~F.}\ \bibnamefont {Carrasquilla}}, \bibinfo
  {author} {\bibfnamefont {Y.~B.}\ \bibnamefont {Kim}},\ and\ \bibinfo {author}
  {\bibfnamefont {H.}~\bibnamefont {Yuen}},\ }\bibfield  {title} {\bibinfo
  {title} {Exploring entanglement and optimization within the {Hamiltonian}
  variational ansatz},\ }\href {https://doi.org/10.1103/PRXQuantum.1.020319}
  {\bibfield  {journal} {\bibinfo  {journal} {PRX Quantum}\ }\textbf {\bibinfo
  {volume} {1}},\ \bibinfo {pages} {020319} (\bibinfo {year}
  {2020})}\BibitemShut {NoStop}%
\bibitem [{\citenamefont {Hadfield}\ \emph {et~al.}(2019)\citenamefont
  {Hadfield}, \citenamefont {Wang}, \citenamefont {O’Gorman}, \citenamefont
  {Rieffel}, \citenamefont {Venturelli},\ and\ \citenamefont
  {Biswas}}]{HaEtAl19}%
  \BibitemOpen
  \bibfield  {author} {\bibinfo {author} {\bibfnamefont {S.}~\bibnamefont
  {Hadfield}}, \bibinfo {author} {\bibfnamefont {Z.}~\bibnamefont {Wang}},
  \bibinfo {author} {\bibfnamefont {B.}~\bibnamefont {O’Gorman}}, \bibinfo
  {author} {\bibfnamefont {E.}~\bibnamefont {Rieffel}}, \bibinfo {author}
  {\bibfnamefont {D.}~\bibnamefont {Venturelli}},\ and\ \bibinfo {author}
  {\bibfnamefont {R.}~\bibnamefont {Biswas}},\ }\bibfield  {title} {\bibinfo
  {title} {From the quantum approximate optimization algorithm to a quantum
  alternating operator ansatz},\ }\href {https://doi.org/10.3390/a12020034}
  {\bibfield  {journal} {\bibinfo  {journal} {Algorithms}\ }\textbf {\bibinfo
  {volume} {12}},\ \bibinfo {pages} {34} (\bibinfo {year} {2019})}\BibitemShut
  {NoStop}%
\bibitem [{\citenamefont {LaRose}\ \emph {et~al.}(2021)\citenamefont {LaRose},
  \citenamefont {Rieffel},\ and\ \citenamefont {Venturelli}}]{LaRose2021}%
  \BibitemOpen
  \bibfield  {author} {\bibinfo {author} {\bibfnamefont {R.}~\bibnamefont
  {LaRose}}, \bibinfo {author} {\bibfnamefont {E.}~\bibnamefont {Rieffel}},\
  and\ \bibinfo {author} {\bibfnamefont {D.}~\bibnamefont {Venturelli}},\
  }\href@noop {} {\bibinfo {title} {Mixer-phaser ans\"atze for quantum
  optimization with hard constraints}} (\bibinfo {year} {2021}),\ \Eprint
  {https://arxiv.org/abs/2107.06651} {arXiv:2107.06651 [quant-ph]} \BibitemShut
  {NoStop}%
\bibitem [{\citenamefont {Majumdar}\ \emph {et~al.}(2021)\citenamefont
  {Majumdar}, \citenamefont {Bhoumik}, \citenamefont {Madan}, \citenamefont
  {Vinayagamurthy}, \citenamefont {Raghunathan},\ and\ \citenamefont
  {Sur-Kolay}}]{MaEtAl21}%
  \BibitemOpen
  \bibfield  {author} {\bibinfo {author} {\bibfnamefont {R.}~\bibnamefont
  {Majumdar}}, \bibinfo {author} {\bibfnamefont {D.}~\bibnamefont {Bhoumik}},
  \bibinfo {author} {\bibfnamefont {D.}~\bibnamefont {Madan}}, \bibinfo
  {author} {\bibfnamefont {D.}~\bibnamefont {Vinayagamurthy}}, \bibinfo
  {author} {\bibfnamefont {S.}~\bibnamefont {Raghunathan}},\ and\ \bibinfo
  {author} {\bibfnamefont {S.}~\bibnamefont {Sur-Kolay}},\ }\href@noop {}
  {\bibinfo {title} {Depth optimized ansatz circuit in {QAOA for Max-Cut}}}
  (\bibinfo {year} {2021}),\ \Eprint {https://arxiv.org/abs/2110.04637}
  {arXiv:2110.04637 [quant-ph]} \BibitemShut {NoStop}%
\bibitem [{\citenamefont {Du}\ \emph {et~al.}(2020)\citenamefont {Du},
  \citenamefont {Huang}, \citenamefont {You}, \citenamefont {Hsieh},\ and\
  \citenamefont {Tao}}]{Du2020}%
  \BibitemOpen
  \bibfield  {author} {\bibinfo {author} {\bibfnamefont {Y.}~\bibnamefont
  {Du}}, \bibinfo {author} {\bibfnamefont {T.}~\bibnamefont {Huang}}, \bibinfo
  {author} {\bibfnamefont {S.}~\bibnamefont {You}}, \bibinfo {author}
  {\bibfnamefont {M.-H.}\ \bibnamefont {Hsieh}},\ and\ \bibinfo {author}
  {\bibfnamefont {D.}~\bibnamefont {Tao}},\ }\href@noop {} {\bibinfo {title}
  {Quantum circuit architecture search: error mitigation and trainability
  enhancement for variational quantum solvers}} (\bibinfo {year} {2020}),\
  \Eprint {https://arxiv.org/abs/2010.10217} {arXiv:2010.10217 [quant-ph]}
  \BibitemShut {NoStop}%
\bibitem [{\citenamefont {Romero}\ \emph {et~al.}(2017)\citenamefont {Romero},
  \citenamefont {Olson},\ and\ \citenamefont {Aspuru-Guzik}}]{RoOlAs17}%
  \BibitemOpen
  \bibfield  {author} {\bibinfo {author} {\bibfnamefont {J.}~\bibnamefont
  {Romero}}, \bibinfo {author} {\bibfnamefont {J.~P.}\ \bibnamefont {Olson}},\
  and\ \bibinfo {author} {\bibfnamefont {A.}~\bibnamefont {Aspuru-Guzik}},\
  }\bibfield  {title} {\bibinfo {title} {Quantum autoencoders for efficient
  compression of quantum data},\ }\href
  {https://doi.org/10.1088/2058-9565/aa8072} {\bibfield  {journal} {\bibinfo
  {journal} {Quantum Sci. Technol.}\ }\textbf {\bibinfo {volume} {2}},\
  \bibinfo {pages} {045001} (\bibinfo {year} {2017})}\BibitemShut {NoStop}%
\bibitem [{\citenamefont {Cao}\ and\ \citenamefont {Wang}(2021)}]{ChWa21}%
  \BibitemOpen
  \bibfield  {author} {\bibinfo {author} {\bibfnamefont {C.}~\bibnamefont
  {Cao}}\ and\ \bibinfo {author} {\bibfnamefont {X.}~\bibnamefont {Wang}},\
  }\bibfield  {title} {\bibinfo {title} {Noise-assisted quantum autoencoder},\
  }\href {https://doi.org/10.1103/PhysRevApplied.15.054012} {\bibfield
  {journal} {\bibinfo  {journal} {Phys. Rev. Applied}\ }\textbf {\bibinfo
  {volume} {15}},\ \bibinfo {pages} {054012} (\bibinfo {year}
  {2021})}\BibitemShut {NoStop}%
\bibitem [{\citenamefont {Foss-Feig}\ \emph
  {et~al.}(2021{\natexlab{a}})\citenamefont {Foss-Feig}, \citenamefont {Hayes},
  \citenamefont {Dreiling}, \citenamefont {Figgatt}, \citenamefont {Gaebler},
  \citenamefont {Moses}, \citenamefont {Pino},\ and\ \citenamefont
  {Potter}}]{FoEtAl20}%
  \BibitemOpen
  \bibfield  {author} {\bibinfo {author} {\bibfnamefont {M.}~\bibnamefont
  {Foss-Feig}}, \bibinfo {author} {\bibfnamefont {D.}~\bibnamefont {Hayes}},
  \bibinfo {author} {\bibfnamefont {J.~M.}\ \bibnamefont {Dreiling}}, \bibinfo
  {author} {\bibfnamefont {C.}~\bibnamefont {Figgatt}}, \bibinfo {author}
  {\bibfnamefont {J.~P.}\ \bibnamefont {Gaebler}}, \bibinfo {author}
  {\bibfnamefont {S.~A.}\ \bibnamefont {Moses}}, \bibinfo {author}
  {\bibfnamefont {J.~M.}\ \bibnamefont {Pino}},\ and\ \bibinfo {author}
  {\bibfnamefont {A.~C.}\ \bibnamefont {Potter}},\ }\bibfield  {title}
  {\bibinfo {title} {Holographic quantum algorithms for simulating correlated
  spin systems},\ }\href {https://doi.org/10.1103/PhysRevResearch.3.033002}
  {\bibfield  {journal} {\bibinfo  {journal} {Phys. Rev. Research}\ }\textbf
  {\bibinfo {volume} {3}},\ \bibinfo {pages} {033002} (\bibinfo {year}
  {2021}{\natexlab{a}})}\BibitemShut {NoStop}%
\bibitem [{\citenamefont {Foss-Feig}\ \emph
  {et~al.}(2021{\natexlab{b}})\citenamefont {Foss-Feig}, \citenamefont
  {Ragole}, \citenamefont {Potter}, \citenamefont {Dreiling}, \citenamefont
  {Figgatt}, \citenamefont {Gaebler}, \citenamefont {Hall}, \citenamefont
  {Moses}, \citenamefont {Pino}, \citenamefont {Spaun}, \citenamefont
  {Neyenhuis},\ and\ \citenamefont {Hayes}}]{FoEtAl21}%
  \BibitemOpen
  \bibfield  {author} {\bibinfo {author} {\bibfnamefont {M.}~\bibnamefont
  {Foss-Feig}}, \bibinfo {author} {\bibfnamefont {S.}~\bibnamefont {Ragole}},
  \bibinfo {author} {\bibfnamefont {A.}~\bibnamefont {Potter}}, \bibinfo
  {author} {\bibfnamefont {J.}~\bibnamefont {Dreiling}}, \bibinfo {author}
  {\bibfnamefont {C.}~\bibnamefont {Figgatt}}, \bibinfo {author} {\bibfnamefont
  {J.}~\bibnamefont {Gaebler}}, \bibinfo {author} {\bibfnamefont
  {A.}~\bibnamefont {Hall}}, \bibinfo {author} {\bibfnamefont {S.}~\bibnamefont
  {Moses}}, \bibinfo {author} {\bibfnamefont {J.}~\bibnamefont {Pino}},
  \bibinfo {author} {\bibfnamefont {B.}~\bibnamefont {Spaun}}, \bibinfo
  {author} {\bibfnamefont {B.}~\bibnamefont {Neyenhuis}},\ and\ \bibinfo
  {author} {\bibfnamefont {D.}~\bibnamefont {Hayes}},\ }\href@noop {} {\bibinfo
  {title} {Entanglement from tensor networks on a trapped-ion {QCCD} quantum
  computer}} (\bibinfo {year} {2021}{\natexlab{b}}),\ \Eprint
  {https://arxiv.org/abs/2104.11235} {arXiv:2104.11235 [quant-ph]} \BibitemShut
  {NoStop}%
\bibitem [{\citenamefont {Chertkov}\ \emph {et~al.}(2021)\citenamefont
  {Chertkov}, \citenamefont {Bohnet}, \citenamefont {Francois}, \citenamefont
  {Gaebler}, \citenamefont {Gresh}, \citenamefont {Hankin}, \citenamefont
  {Lee}, \citenamefont {Tobey}, \citenamefont {Hayes}, \citenamefont
  {Neyenhuis}, \citenamefont {Stutz}, \citenamefont {Potter},\ and\
  \citenamefont {Foss-Feig}}]{ChEtAl21}%
  \BibitemOpen
  \bibfield  {author} {\bibinfo {author} {\bibfnamefont {E.}~\bibnamefont
  {Chertkov}}, \bibinfo {author} {\bibfnamefont {J.}~\bibnamefont {Bohnet}},
  \bibinfo {author} {\bibfnamefont {D.}~\bibnamefont {Francois}}, \bibinfo
  {author} {\bibfnamefont {J.}~\bibnamefont {Gaebler}}, \bibinfo {author}
  {\bibfnamefont {D.}~\bibnamefont {Gresh}}, \bibinfo {author} {\bibfnamefont
  {A.}~\bibnamefont {Hankin}}, \bibinfo {author} {\bibfnamefont
  {K.}~\bibnamefont {Lee}}, \bibinfo {author} {\bibfnamefont {R.}~\bibnamefont
  {Tobey}}, \bibinfo {author} {\bibfnamefont {D.}~\bibnamefont {Hayes}},
  \bibinfo {author} {\bibfnamefont {B.}~\bibnamefont {Neyenhuis}}, \bibinfo
  {author} {\bibfnamefont {R.}~\bibnamefont {Stutz}}, \bibinfo {author}
  {\bibfnamefont {A.~C.}\ \bibnamefont {Potter}},\ and\ \bibinfo {author}
  {\bibfnamefont {M.}~\bibnamefont {Foss-Feig}},\ }\href@noop {} {\bibinfo
  {title} {Holographic dynamics simulations with a trapped ion quantum
  computer}} (\bibinfo {year} {2021}),\ \Eprint
  {https://arxiv.org/abs/2105.09324} {arXiv:2105.09324 [quant-ph]} \BibitemShut
  {NoStop}%
\bibitem [{\citenamefont {Zhang}\ \emph {et~al.}(2021)\citenamefont {Zhang},
  \citenamefont {Wan}, \citenamefont {Lee}, \citenamefont {Hsieh},
  \citenamefont {Zhang},\ and\ \citenamefont {Yao}}]{ZhEtAl21}%
  \BibitemOpen
  \bibfield  {author} {\bibinfo {author} {\bibfnamefont {S.-X.}\ \bibnamefont
  {Zhang}}, \bibinfo {author} {\bibfnamefont {Z.-Q.}\ \bibnamefont {Wan}},
  \bibinfo {author} {\bibfnamefont {C.-K.}\ \bibnamefont {Lee}}, \bibinfo
  {author} {\bibfnamefont {C.-Y.}\ \bibnamefont {Hsieh}}, \bibinfo {author}
  {\bibfnamefont {S.}~\bibnamefont {Zhang}},\ and\ \bibinfo {author}
  {\bibfnamefont {H.}~\bibnamefont {Yao}},\ }\href@noop {} {\bibinfo {title}
  {Variational quantum-neural hybrid eigensolver}} (\bibinfo {year} {2021}),\
  \Eprint {https://arxiv.org/abs/2106.05105} {arXiv:2106.05105 [quant-ph]}
  \BibitemShut {NoStop}%
\bibitem [{\citenamefont {Ostaszewski}\ \emph
  {et~al.}(2021{\natexlab{a}})\citenamefont {Ostaszewski}, \citenamefont
  {Grant},\ and\ \citenamefont {Benedetti}}]{Ostaszewski_Benedetti_2021}%
  \BibitemOpen
  \bibfield  {author} {\bibinfo {author} {\bibfnamefont {M.}~\bibnamefont
  {Ostaszewski}}, \bibinfo {author} {\bibfnamefont {E.}~\bibnamefont {Grant}},\
  and\ \bibinfo {author} {\bibfnamefont {M.}~\bibnamefont {Benedetti}},\
  }\bibfield  {title} {\bibinfo {title} {Structure optimization for
  parameterized quantum circuits},\ }\href
  {https://doi.org/10.22331/q-2021-01-28-391} {\bibfield  {journal} {\bibinfo
  {journal} {{Quantum}}\ }\textbf {\bibinfo {volume} {5}},\ \bibinfo {pages}
  {391} (\bibinfo {year} {2021}{\natexlab{a}})}\BibitemShut {NoStop}%
\bibitem [{\citenamefont {Ostaszewski}\ \emph
  {et~al.}(2021{\natexlab{b}})\citenamefont {Ostaszewski}, \citenamefont
  {Trenkwalder}, \citenamefont {Masarczyk}, \citenamefont {Scerri},\ and\
  \citenamefont {Dunjko}}]{Ostaszewski2021}%
  \BibitemOpen
  \bibfield  {author} {\bibinfo {author} {\bibfnamefont {M.}~\bibnamefont
  {Ostaszewski}}, \bibinfo {author} {\bibfnamefont {L.~M.}\ \bibnamefont
  {Trenkwalder}}, \bibinfo {author} {\bibfnamefont {W.}~\bibnamefont
  {Masarczyk}}, \bibinfo {author} {\bibfnamefont {E.}~\bibnamefont {Scerri}},\
  and\ \bibinfo {author} {\bibfnamefont {V.}~\bibnamefont {Dunjko}},\
  }\href@noop {} {\bibinfo {title} {Reinforcement learning for optimization of
  variational quantum circuit architectures}} (\bibinfo {year}
  {2021}{\natexlab{b}}),\ \Eprint {https://arxiv.org/abs/2103.16089}
  {arXiv:2103.16089 [quant-ph]} \BibitemShut {NoStop}%
\bibitem [{\citenamefont {Grimsley}\ \emph {et~al.}(2019)\citenamefont
  {Grimsley}, \citenamefont {Economou}, \citenamefont {Barnes},\ and\
  \citenamefont {Mayhall}}]{GrEtAl19}%
  \BibitemOpen
  \bibfield  {author} {\bibinfo {author} {\bibfnamefont {H.~R.}\ \bibnamefont
  {Grimsley}}, \bibinfo {author} {\bibfnamefont {S.~E.}\ \bibnamefont
  {Economou}}, \bibinfo {author} {\bibfnamefont {E.}~\bibnamefont {Barnes}},\
  and\ \bibinfo {author} {\bibfnamefont {N.~J.}\ \bibnamefont {Mayhall}},\
  }\bibfield  {title} {\bibinfo {title} {An adaptive variational algorithm for
  exact molecular simulations on a quantum computer},\ }\href
  {https://doi.org/10.1038/s41467-019-10988-2} {\bibfield  {journal} {\bibinfo
  {journal} {Nat. Commun.}\ }\textbf {\bibinfo {volume} {10}},\ \bibinfo
  {pages} {3007} (\bibinfo {year} {2019})}\BibitemShut {NoStop}%
\bibitem [{\citenamefont {Zhu}\ \emph {et~al.}(2020)\citenamefont {Zhu},
  \citenamefont {Tang}, \citenamefont {Barron}, \citenamefont
  {Calderon-Vargas}, \citenamefont {Mayhall}, \citenamefont {Barnes},\ and\
  \citenamefont {Economou}}]{ZhEtAl20}%
  \BibitemOpen
  \bibfield  {author} {\bibinfo {author} {\bibfnamefont {L.}~\bibnamefont
  {Zhu}}, \bibinfo {author} {\bibfnamefont {H.~L.}\ \bibnamefont {Tang}},
  \bibinfo {author} {\bibfnamefont {G.~S.}\ \bibnamefont {Barron}}, \bibinfo
  {author} {\bibfnamefont {F.~A.}\ \bibnamefont {Calderon-Vargas}}, \bibinfo
  {author} {\bibfnamefont {N.~J.}\ \bibnamefont {Mayhall}}, \bibinfo {author}
  {\bibfnamefont {E.}~\bibnamefont {Barnes}},\ and\ \bibinfo {author}
  {\bibfnamefont {S.~E.}\ \bibnamefont {Economou}},\ }\href@noop {} {\bibinfo
  {title} {An adaptive quantum approximate optimization algorithm for solving
  combinatorial problems on a quantum computer}} (\bibinfo {year} {2020}),\
  \Eprint {https://arxiv.org/abs/2005.10258} {arXiv:2005.10258 [quant-ph]}
  \BibitemShut {NoStop}%
\bibitem [{\citenamefont {Skolik}\ \emph {et~al.}(2021)\citenamefont {Skolik},
  \citenamefont {McClean}, \citenamefont {Mohseni}, \citenamefont {van~der
  Smagt},\ and\ \citenamefont {Leib}}]{SkEtAl21}%
  \BibitemOpen
  \bibfield  {author} {\bibinfo {author} {\bibfnamefont {A.}~\bibnamefont
  {Skolik}}, \bibinfo {author} {\bibfnamefont {J.~R.}\ \bibnamefont {McClean}},
  \bibinfo {author} {\bibfnamefont {M.}~\bibnamefont {Mohseni}}, \bibinfo
  {author} {\bibfnamefont {P.}~\bibnamefont {van~der Smagt}},\ and\ \bibinfo
  {author} {\bibfnamefont {M.}~\bibnamefont {Leib}},\ }\bibfield  {title}
  {\bibinfo {title} {Layerwise learning for quantum neural networks},\ }\href
  {https://doi.org/10.1007/s42484-020-00036-4} {\bibfield  {journal} {\bibinfo
  {journal} {Quantum Mach. Intell.}\ }\textbf {\bibinfo {volume} {3}},\
  \bibinfo {pages} {5} (\bibinfo {year} {2021})}\BibitemShut {NoStop}%
\bibitem [{\citenamefont {Liu}\ \emph {et~al.}(2021)\citenamefont {Liu},
  \citenamefont {Angone}, \citenamefont {Shaydulin}, \citenamefont {Safro},
  \citenamefont {Alexeev},\ and\ \citenamefont {Cincio}}]{Liu2021}%
  \BibitemOpen
  \bibfield  {author} {\bibinfo {author} {\bibfnamefont {X.}~\bibnamefont
  {Liu}}, \bibinfo {author} {\bibfnamefont {A.}~\bibnamefont {Angone}},
  \bibinfo {author} {\bibfnamefont {R.}~\bibnamefont {Shaydulin}}, \bibinfo
  {author} {\bibfnamefont {I.}~\bibnamefont {Safro}}, \bibinfo {author}
  {\bibfnamefont {Y.}~\bibnamefont {Alexeev}},\ and\ \bibinfo {author}
  {\bibfnamefont {L.}~\bibnamefont {Cincio}},\ }\href@noop {} {\bibinfo {title}
  {{Layer VQE: A} variational approach for combinatorial optimization on noisy
  quantum computers}} (\bibinfo {year} {2021}),\ \Eprint
  {https://arxiv.org/abs/2102.05566} {arXiv:2102.05566 [quant-ph]} \BibitemShut
  {NoStop}%
\bibitem [{\citenamefont {Sweke}\ \emph {et~al.}(2020)\citenamefont {Sweke},
  \citenamefont {Wilde}, \citenamefont {Meyer}, \citenamefont {Schuld},
  \citenamefont {Faehrmann}, \citenamefont {Meynard-Piganeau},\ and\
  \citenamefont {Eisert}}]{Sweke2020}%
  \BibitemOpen
  \bibfield  {author} {\bibinfo {author} {\bibfnamefont {R.}~\bibnamefont
  {Sweke}}, \bibinfo {author} {\bibfnamefont {F.}~\bibnamefont {Wilde}},
  \bibinfo {author} {\bibfnamefont {J.}~\bibnamefont {Meyer}}, \bibinfo
  {author} {\bibfnamefont {M.}~\bibnamefont {Schuld}}, \bibinfo {author}
  {\bibfnamefont {P.~K.}\ \bibnamefont {Faehrmann}}, \bibinfo {author}
  {\bibfnamefont {B.}~\bibnamefont {Meynard-Piganeau}},\ and\ \bibinfo {author}
  {\bibfnamefont {J.}~\bibnamefont {Eisert}},\ }\bibfield  {title} {\bibinfo
  {title} {Stochastic gradient descent for hybrid quantum-classical
  optimization},\ }\href {https://doi.org/10.22331/q-2020-08-31-314} {\bibfield
   {journal} {\bibinfo  {journal} {{Quantum}}\ }\textbf {\bibinfo {volume}
  {4}},\ \bibinfo {pages} {314} (\bibinfo {year} {2020})}\BibitemShut {NoStop}%
\bibitem [{\citenamefont {Kirkpatrick}\ \emph {et~al.}(1983)\citenamefont
  {Kirkpatrick}, \citenamefont {Gelatt},\ and\ \citenamefont
  {Vecchi}}]{KiGeVe83}%
  \BibitemOpen
  \bibfield  {author} {\bibinfo {author} {\bibfnamefont {S.}~\bibnamefont
  {Kirkpatrick}}, \bibinfo {author} {\bibfnamefont {C.~D.}\ \bibnamefont
  {Gelatt}},\ and\ \bibinfo {author} {\bibfnamefont {M.~P.}\ \bibnamefont
  {Vecchi}},\ }\bibfield  {title} {\bibinfo {title} {Optimization by simulated
  annealing},\ }\href {https://doi.org/10.1126/science.220.4598.671} {\bibfield
   {journal} {\bibinfo  {journal} {Science}\ }\textbf {\bibinfo {volume}
  {220}},\ \bibinfo {pages} {671} (\bibinfo {year} {1983})}\BibitemShut
  {NoStop}%
\bibitem [{\citenamefont {Rothman}(1985)}]{Ro85}%
  \BibitemOpen
  \bibfield  {author} {\bibinfo {author} {\bibfnamefont {D.~H.}\ \bibnamefont
  {Rothman}},\ }\href
  {http://sepwww.stanford.edu/data/media/public/oldreports/sep45/} {\emph
  {\bibinfo {title} {Large near-surface anomalies, seismic...}}}\ (\bibinfo
  {publisher} {Ph.D. thesis, Stanford University},\ \bibinfo {year}
  {1985})\BibitemShut {NoStop}%
\bibitem [{\citenamefont {Kyriienko}(2020)}]{Kyriienko2020}%
  \BibitemOpen
  \bibfield  {author} {\bibinfo {author} {\bibfnamefont {O.}~\bibnamefont
  {Kyriienko}},\ }\bibfield  {title} {\bibinfo {title} {Quantum inverse
  iteration algorithm for programmable quantum simulators},\ }\href
  {https://doi.org/10.1038/s41534-019-0239-7} {\bibfield  {journal} {\bibinfo
  {journal} {npj Quantum Inf.}\ }\textbf {\bibinfo {volume} {6}},\ \bibinfo
  {pages} {7} (\bibinfo {year} {2020})}\BibitemShut {NoStop}%
\bibitem [{\citenamefont {Zeng}\ \emph {et~al.}(2021)\citenamefont {Zeng},
  \citenamefont {Sun},\ and\ \citenamefont {Yuan}}]{ZeSuYu21}%
  \BibitemOpen
  \bibfield  {author} {\bibinfo {author} {\bibfnamefont {P.}~\bibnamefont
  {Zeng}}, \bibinfo {author} {\bibfnamefont {J.}~\bibnamefont {Sun}},\ and\
  \bibinfo {author} {\bibfnamefont {X.}~\bibnamefont {Yuan}},\ }\href@noop {}
  {\bibinfo {title} {Universal quantum algorithmic cooling on a quantum
  computer}} (\bibinfo {year} {2021}),\ \Eprint
  {https://arxiv.org/abs/2109.15304} {arXiv:2109.15304 [quant-ph]} \BibitemShut
  {NoStop}%
\bibitem [{\citenamefont {Alcazar}\ and\ \citenamefont
  {Perdomo-Ortiz}(2021)}]{AlPe21}%
  \BibitemOpen
  \bibfield  {author} {\bibinfo {author} {\bibfnamefont {J.}~\bibnamefont
  {Alcazar}}\ and\ \bibinfo {author} {\bibfnamefont {A.}~\bibnamefont
  {Perdomo-Ortiz}},\ }\href@noop {} {\bibinfo {title} {Enhancing combinatorial
  optimization with quantum generative models}} (\bibinfo {year} {2021}),\
  \Eprint {https://arxiv.org/abs/2101.06250} {arXiv:2101.06250 [quant-ph]}
  \BibitemShut {NoStop}%
\bibitem [{\citenamefont {Garc{\'{i}}a-Ripoll}(2021)}]{Ga21}%
  \BibitemOpen
  \bibfield  {author} {\bibinfo {author} {\bibfnamefont {J.~J.}\ \bibnamefont
  {Garc{\'{i}}a-Ripoll}},\ }\bibfield  {title} {\bibinfo {title}
  {Quantum-inspired algorithms for multivariate analysis: from interpolation to
  partial differential equations},\ }\href
  {https://doi.org/10.22331/q-2021-04-15-431} {\bibfield  {journal} {\bibinfo
  {journal} {{Quantum}}\ }\textbf {\bibinfo {volume} {5}},\ \bibinfo {pages}
  {431} (\bibinfo {year} {2021})}\BibitemShut {NoStop}%
\bibitem [{\citenamefont {Patti}\ \emph {et~al.}(2022)\citenamefont {Patti},
  \citenamefont {Kossaifi}, \citenamefont {Anandkumar},\ and\ \citenamefont
  {Yelin}}]{Patti2021}%
  \BibitemOpen
  \bibfield  {author} {\bibinfo {author} {\bibfnamefont {T.~L.}\ \bibnamefont
  {Patti}}, \bibinfo {author} {\bibfnamefont {J.}~\bibnamefont {Kossaifi}},
  \bibinfo {author} {\bibfnamefont {A.}~\bibnamefont {Anandkumar}},\ and\
  \bibinfo {author} {\bibfnamefont {S.~F.}\ \bibnamefont {Yelin}},\ }\href@noop
  {} {\bibinfo {title} {Variational quantum optimization with multi-basis
  encodings}} (\bibinfo {year} {2022}),\ \Eprint
  {https://arxiv.org/abs/2106.13304} {arXiv:2106.13304 [quant-ph]} \BibitemShut
  {NoStop}%
\bibitem [{\citenamefont {Lubasch}\ \emph {et~al.}(2016)\citenamefont
  {Lubasch}, \citenamefont {Fuks}, \citenamefont {Appel}, \citenamefont
  {Rubio}, \citenamefont {Cirac},\ and\ \citenamefont
  {Ba{\~{n}}uls}}]{LuEtAl16}%
  \BibitemOpen
  \bibfield  {author} {\bibinfo {author} {\bibfnamefont {M.}~\bibnamefont
  {Lubasch}}, \bibinfo {author} {\bibfnamefont {J.~I.}\ \bibnamefont {Fuks}},
  \bibinfo {author} {\bibfnamefont {H.}~\bibnamefont {Appel}}, \bibinfo
  {author} {\bibfnamefont {A.}~\bibnamefont {Rubio}}, \bibinfo {author}
  {\bibfnamefont {J.~I.}\ \bibnamefont {Cirac}},\ and\ \bibinfo {author}
  {\bibfnamefont {M.-C.}\ \bibnamefont {Ba{\~{n}}uls}},\ }\bibfield  {title}
  {\bibinfo {title} {Systematic construction of density functionals based on
  matrix product state computations},\ }\href
  {https://doi.org/10.1088/1367-2630/18/8/083039} {\bibfield  {journal}
  {\bibinfo  {journal} {New J. of Phys.}\ }\textbf {\bibinfo {volume} {18}},\
  \bibinfo {pages} {083039} (\bibinfo {year} {2016})}\BibitemShut {NoStop}%
\bibitem [{\citenamefont {Lubasch}\ \emph {et~al.}(2018)\citenamefont
  {Lubasch}, \citenamefont {Moinier},\ and\ \citenamefont {Jaksch}}]{LuMoJa18}%
  \BibitemOpen
  \bibfield  {author} {\bibinfo {author} {\bibfnamefont {M.}~\bibnamefont
  {Lubasch}}, \bibinfo {author} {\bibfnamefont {P.}~\bibnamefont {Moinier}},\
  and\ \bibinfo {author} {\bibfnamefont {D.}~\bibnamefont {Jaksch}},\
  }\bibfield  {title} {\bibinfo {title} {Multigrid renormalization},\ }\href
  {https://doi.org/https://doi.org/10.1016/j.jcp.2018.06.065} {\bibfield
  {journal} {\bibinfo  {journal} {J. Comput. Phys.}\ }\textbf {\bibinfo
  {volume} {372}},\ \bibinfo {pages} {587} (\bibinfo {year}
  {2018})}\BibitemShut {NoStop}%
\bibitem [{\citenamefont {Orús}(2014)}]{Or14}%
  \BibitemOpen
  \bibfield  {author} {\bibinfo {author} {\bibfnamefont {R.}~\bibnamefont
  {Orús}},\ }\bibfield  {title} {\bibinfo {title} {A practical introduction to
  tensor networks: Matrix product states and projected entangled pair states},\
  }\href {https://doi.org/https://doi.org/10.1016/j.aop.2014.06.013} {\bibfield
   {journal} {\bibinfo  {journal} {Ann. of Phys.}\ }\textbf {\bibinfo {volume}
  {349}},\ \bibinfo {pages} {117} (\bibinfo {year} {2014})}\BibitemShut
  {NoStop}%
\bibitem [{\citenamefont {Cirac}\ \emph {et~al.}(2021)\citenamefont {Cirac},
  \citenamefont {P\'erez-Garc\'{\i}a}, \citenamefont {Schuch},\ and\
  \citenamefont {Verstraete}}]{CiEtAl20}%
  \BibitemOpen
  \bibfield  {author} {\bibinfo {author} {\bibfnamefont {J.~I.}\ \bibnamefont
  {Cirac}}, \bibinfo {author} {\bibfnamefont {D.}~\bibnamefont
  {P\'erez-Garc\'{\i}a}}, \bibinfo {author} {\bibfnamefont {N.}~\bibnamefont
  {Schuch}},\ and\ \bibinfo {author} {\bibfnamefont {F.}~\bibnamefont
  {Verstraete}},\ }\bibfield  {title} {\bibinfo {title} {Matrix product states
  and projected entangled pair states: {Concepts}, symmetries, theorems},\
  }\href {https://doi.org/10.1103/RevModPhys.93.045003} {\bibfield  {journal}
  {\bibinfo  {journal} {Rev. Mod. Phys.}\ }\textbf {\bibinfo {volume} {93}},\
  \bibinfo {pages} {045003} (\bibinfo {year} {2021})}\BibitemShut {NoStop}%
\end{thebibliography}%

\end{document}